\documentclass[fleqn,usenatbib]{mnras}

\usepackage{newtxtext,newtxmath}

\usepackage[T1]{fontenc}
\usepackage{ae,aecompl}

\usepackage{graphicx}	
\usepackage{amsmath}	

\title[Exoplanet variability with vAPP coronagraphs]{Measuring the variability of directly imaged exoplanets using vector Apodizing Phase Plates combined with ground-based differential spectrophotometry} 

\author[B. J. Sutlieff et al.]{Ben J. Sutlieff,$^{1,2}$\thanks{E-mail: b.j.sutlieff@uva.nl}
Jayne L. Birkby,$^{3,1}$
Jordan M. Stone,$^{4}$
David S. Doelman,$^{2}$
\newauthor Matthew A. Kenworthy,$^{2}$
Vatsal Panwar,$^{1,5}$
Alexander J. Bohn,$^{2}$
Steve Ertel,$^{6,7}$
\newauthor Frans Snik,$^{2}$
Charles E. Woodward,$^{8}$
Andrew J. Skemer,$^{9}$
Jarron M. Leisenring,$^{7}$
\newauthor Klaus G. Strassmeier,$^{10}$
and David Charbonneau$^{11}$
\\
$^{1}$Anton Pannekoek Institute for Astronomy, University of Amsterdam, Science Park 904, 1098 XH Amsterdam, The Netherlands\\
$^{2}$Leiden Observatory, Leiden University, P.O. Box 9513, 2300 RA Leiden, The Netherlands\\
$^{3}$Astrophysics, University of Oxford, Denys Wilkinson Building, Keble Road, Oxford, OX1 3RH, United Kingdom\\
$^{4}$Naval Research Laboratory, Remote Sensing Division, 4555 Overlook Ave. SW, Washington, DC 20375, USA\\
$^{5}$Department of Physics, University of Warwick, Coventry
West Midlands, CV4 7AL, United Kingdom\\
$^{6}$Large Binocular Telescope Observatory, University of Arizona, 933 North Cherry Avenue, Tucson, AZ 85721, USA\\
$^{7}$Steward Observatory and Department of Astronomy, University of Arizona, 933 N. Cherry Ave., Tucson, AZ 85721, USA\\
$^{8}$Minnesota Institute for Astrophysics, University of Minnesota, 116 Church Street SE, Minneapolis, MN 55455, USA\\
$^{9}$Department of Astronomy and Astrophysics, University of California, Santa Cruz, 1156 High St, Santa Cruz, CA 95064, USA\\
$^{10}$Leibniz-Institute for Astrophysics Potsdam (AIP), An der Sternwarte 16, 14482 Potsdam, Germany\\
$^{11}$Center for Astrophysics \textbar~Harvard \& Smithsonian, 60 Garden Street, Cambridge, MA 02138, USA
}
\date{Accepted XXX. Received YYY; in original form ZZZ}

\pubyear{2022}

\begin{document}
\label{firstpage}
\pagerange{\pageref{firstpage}--\pageref{lastpage}}
\maketitle

\begin{abstract}
Clouds and other features in exoplanet and brown dwarf atmospheres cause variations in brightness as they rotate in and out of view. Ground-based instruments reach the high contrasts and small inner working angles needed to monitor these faint companions, but their small fields-of-view lack simultaneous photometric references to correct for non-astrophysical variations. We present a novel approach for making ground-based light curves of directly imaged companions using high-cadence differential spectrophotometric monitoring, where the simultaneous reference is provided by a double-grating 360\textdegree{} vector Apodizing Phase Plate (dgvAPP360) coronagraph. The dgvAPP360 enables high-contrast companion detections without blocking the host star, allowing it to be used as a simultaneous reference. To further reduce systematic noise, we emulate exoplanet transmission spectroscopy, where the light is spectrally-dispersed and then recombined into white-light flux. We do this by combining the dgvAPP360 with the infrared ALES integral field spectrograph on the Large Binocular Telescope Interferometer. To demonstrate, we observed the red companion HD~1160~B (separation $\sim$780 mas) for one night, and detect $8.8\%$ semi-amplitude sinusoidal variability with a $\sim$3.24~h period in its detrended white-light curve. We achieve the greatest precision in ground-based high-contrast imaging light curves of sub-arcsecond companions to date, reaching $3.7\%$ precision per 18-minute bin. Individual wavelength channels spanning 3.59-3.99~\textmu m further show tentative evidence of increasing variability with wavelength. We find no evidence yet of a systematic noise floor, hence additional observations can further improve the precision. This is therefore a promising avenue for future work aiming to map storms or find transiting exomoons around giant exoplanets.

\end{abstract}

\section{Introduction}\label{intro}
\begin{figure*}
	\includegraphics[scale=0.3]{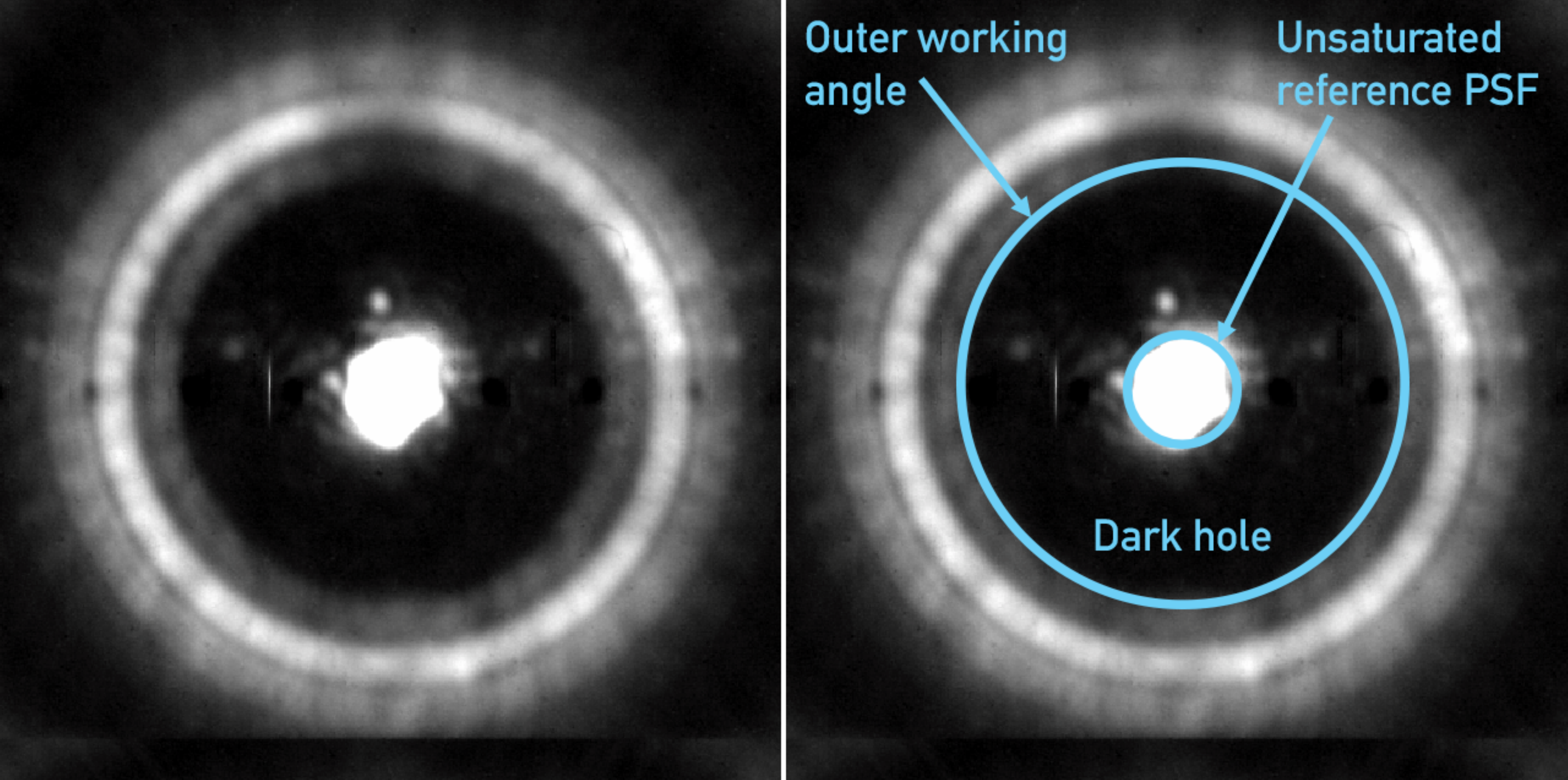}
    \caption{An on-sky example image of a $K\sim7$ mag target observed with the dgvAPP360 coronagraph on the Large Binocular Telescope (LBT), copied and annotated on the right. The dgvAPP360 produces a dark hole of deep flux suppression, seen here surrounding the target star, allowing high-contrast companions to be detected in this region. The star itself remains unsaturated in the middle, allowing it to be used as a simultaneous reference PSF when making a differential light curve for a companion. The outer edge of the dark hole lies beyond the field of view in the LBT observations used in this work.}
    \label{fig:vapp360_example}
\end{figure*}
Planets do not have a homogeneous appearance. When we look at the planets in our own solar system, we see distinct cloud structures and giant storms that show great diversity in size, shape, lifetime, and brightness \citep[e.g.][]{2016ApJ...817..162S, 2021PSJ.....2...47S, 2016AJ....152..142S, 2019AJ....157...89G, 2022ApJS..263...15C}. These features rotate in and out of view throughout the planet's rotation period, modulating its overall brightness and thus allowing us to map out its atmosphere \citep[e.g.][]{2013ApJ...762...47K, 2015ApJ...814...65K, 2016ApJ...825...90K, 2016Icar..278..128F, 2017Sci...357..683A, 2022ApJ...933..163P}. Beyond the solar system, variations in the light curves of stars deliver information on the distribution of features such as star spots \citep{2002AN....323..333B, 2007MNRAS.375..567J, 2009A&A...508.1313F, 2009A&ARv..17..251S, 2010ApJ...718..502M, 2012MNRAS.427.3358G, 2012JAVSO..40..448H, 2013A&A...557L..10N, 2021ApJ...920..132P, 2021MNRAS.502.1299T}. By measuring the photometric variability of exoplanets and brown dwarfs in the same way, we can gain not only an insight into their visual appearance, but also key information on the distribution of condensate clouds that strongly affect the infrared spectra of directly imaged companions, allowing degeneracies between atmospheric models to be broken \citep[e.g.][]{2016ApJ...826....8Y, 2017AJ....154...10R, 2018ApJ...854..172C, 2019AJ....157..128Z, 2020RAA....20...99Z, 2021MNRAS.502..678T, 2021AJ....161....5W}. Space-based photometric monitoring with the Hubble Space Telescope (HST) has already shown that giant planetary-mass and brown dwarf companions do exhibit such variability, at a range of amplitudes and periods \citep{2013ApJ...768..121A, 2014ApJ...782...77B, 2015ApJ...798..127B, 2015ApJ...812..163B, 2016ApJ...818..176Z, 2020AJ....159..140Z, 2018AJ....155...11M, 2019ApJ...875L..15M, 2019ApJ...883..181M, 2020ApJ...893L..30B, 2020AJ....159..125L}. These results are in good agreement with observations of isolated brown dwarfs and giant exoplanet analogues \citep{2018AJ....155...95B, 2018MNRAS.474.1041V, 2020AJ....160...38V,  2020ApJ...903...15L, 2022ApJ...934..178A}, including a large Spitzer survey by \citet{2015ApJ...799..154M} who found that photospheric spots causing $\geq$0.2\% variability at 3-5~\textmu m are ubiquitous. Several studies have identified objects with much stronger variability, at the $>$10\% level, with some even varying with peak-to-peak amplitudes as high as 26\% \citep[e.g.][]{2012ApJ...750..105R, 2014A&A...566A.111W, 2015ApJ...813L..23B, 2019A&A...629A.145E, 2020ApJ...893L..30B}. \citet{2022ApJ...924...68V} further found that young, low-mass brown dwarfs with similar colours and spectra to directly imaged exoplanetary companions are highly likely to display variability in the L2-T4 spectral type range, with an enhancement in maximum amplitudes compared to field dwarfs.

The rotation periods of brown dwarf and planetary-mass companions are consistent with those of the isolated low-mass brown dwarf population, suggesting that they may share similar angular momentum histories \citep{2018NatAs...2..138B, 2022ApJ...924...68V}. These periods are generally short, ranging from $\sim$1~hour to $\gtrsim$20~hours \citep[e.g.][]{2014ApJ...793...75R, 2015ApJ...799..154M, 2021ApJ...906...64A, 2021AJ....161..224T}, within the range expected when evolutionary models and the age- and mass-dependent breakup velocities are considered \citep{2016ApJ...830..141L, 2020AJ....160...38V}. These periods, derived from photometric measurements, are complementary to measurements of companion spin obtained from their spectra \citep{2014Natur.509...63S, 2016A&A...593A..74S, 2018NatAs...2..138B, 2020ApJ...905...37B, 2020AJ....159...97X, 2021AJ....162..148W}. When combined, rotation period and spin measurements can be used to constrain companion obliquities \citep{2020AJ....159..181B, 2021AJ....162..217B}.
However, the population of directly imaged companions accessible to space-based facilities such as HST remains small as most companions lie at close angular separations within the inner working angles of these facilities.

Equipped with coronagraphs and extreme adaptive optics (AO) systems operating in the infrared, large ground-based observatories have the resolution and photon collecting power needed to overcome the glare of the host star and reach the high contrasts and close angular separations of substellar companions currently inaccessible to space telescopes \citep{2016PASP..128j2001B, 2021exbi.book....5H, 2022arXiv220505696C}. Although this provides the opportunity for variability studies of such companions, precise photometric monitoring is difficult as the companion light curve is contaminated by variability arising from Earth's atmosphere and other systematics. Therefore, a simultaneous, unsaturated photometric reference is required to remove this contaminant variability from the companion light curve. For non-coronagraphic, ground-based observations of isolated brown dwarfs and planetary-mass objects, nonvariable stars present in the field of view have often been used as photometric references to enable many successful measurements of variability \citep[e.g.][]{2002ApJ...577..433G, 2013ApJ...778L..10B, 2015ApJ...813L..23B, 2013ApJ...767...61G, 2014ApJ...793...75R, 2014A&A...566A.111W, 2017AJ....154..138N, 2019A&A...629A.145E, 2019MNRAS.483..480V, 2021AJ....162..179M, 2022AJ....164...65M}. However, the typically narrow field of view of ground-based coronagraphic imagers generally precludes the use of field stars as photometric references for observations of companions, and widely used focal-plane coronagraphs block the host star to enable the detection of the companion \citep{2005ApJ...618L.161S, 2012SPIE.8442E..04M, 2018SPIE10698E..2SR}.

One solution to this problem is to use off-axis satellite Point Spread Functions (PSFs), or satellite spots, which can act as simultaneous photometric references even when a host star is blocked by a coronagraph \citep{2006ApJ...647..612M, 2006ApJ...647..620S}. Satellite spots can be created by adding a periodic modulation to the deformable mirror of an AO-equipped telescope or by placing a square grid in the pupil plane to produce spots through diffraction of starlight \citep{2013aoel.confE..63L, 2014SPIE.9147E..55W, 2015ApJ...813L..24J}. The former approach has been used by \citet{2016ApJ...820...40A}, \citet{2021MNRAS.503..743B}, and \citet{2022AJ....164..143W} for observations of the multi-planet HR~8799 system \citep{2008Sci...322.1348M, 2010Natur.468.1080M}. The first two of these studies observed HR~8799 with the Spectro-Polarimetric High-contrast imager for Exoplanets REsearch \citep[SPHERE;][]{2019A&A...631A.155B} instrument at the Very Large Telescope (VLT), and the latter used the CHARIS integral field spectrograph (IFS) in combination with the Subaru Coronagraphic Extreme Adaptive Optics instrument at the Subaru Telescope \citep{2015PASP..127..890J, 2017SPIE10400E..16G}. \citet{2021MNRAS.503..743B} used the satellite spots with a broadband-H filter to successfully constrain their sensitivity to variability to amplitudes $>$5\% for HR~8799b for periods $<$10 hours, and amplitudes $>$25\% for HR~8799c for similar periods, noting that the observed amplitude of any variability would be muted by the likely pole-on viewing angle of these planets \citep{2017ApJ...842...78V, 2018AJ....156..192W, 2019AJ....158..200R}. They also rule out non-shared variability between HR~8799b and HR~8799c at the $<$10-20\% level over a 4-5 hour timescale by using one planet as a photometric reference for the other. Using a spectrophotometric approach, \citet{2022AJ....164..143W} further constrained the variability amplitudes of HR~8799c to the 10\% level, and HR~8799d to the 30\% level, and found that there was no significant variability in the planet's colours. However, all three studies found that satellite spots are anti-correlated with each other and can demonstrate individual variations of their own, potentially setting a limit to the precision that can be achieved with this technique (although \citet{2021MNRAS.503..743B} and \citet{2022AJ....164..143W} note that the satellite spot light curves can be flat in their most stable epochs).
\subsection{Ground-based differential spectrophotometry}\label{intro_diff_spec}
In this paper we present a novel, alternative ground-based approach for constructing light curves of high-contrast companions directly through the technique of differential spectrophotometric monitoring, akin to that used highly successfully to study exoplanet transmission spectra \citep[e.g.][]{2014Natur.505...69K, 2014AJ....147..161S, 2020MNRAS.497.5155W, 2021A&A...646A..94A, 2022MNRAS.510.3236P, 2022MNRAS.515.5018P}. We use a double-grating 360\textdegree{} vector Apodizing Phase Plate (dgvAPP360) coronagraph \citep{2017SPIE10400E..0UD, 2020PASP..132d5002D, 2021ApOpt..60D..52D, 2020AJ....159..252W}, which enables high-contrast companions to be detected without blocking the host star, hence leaving an unsaturated image of the host star available for use as a simultaneous reference. The more widely used grating vector Apodizing Phase Plate (gvAPP) coronagraph adjusts the phase of the incoming wavefront to modify the PSFs of all objects in the field of view, creating two images of the target star each with a 180\textdegree{} D-shaped `dark hole', a region of deep flux suppression in which high-contrast companions can be observed, on opposing sides \citep{2012SPIE.8450E..0MS, 2014OExpr..2230287O, 2017ApJ...834..175O, 2020SPIE11448E..3WB, 2021ApOpt..60D..52D, 2021MNRAS.506.3224S}. The dgvAPP360 instead creates a 360\textdegree{} dark hole surrounding each of the two images of the target star, and then uses an additional grating to overlap the images to produce a single image of the star \citep{2022AJ....163..217D}. An example image obtained with the dgvAPP360 is shown in Figure \ref{fig:vapp360_example}, with the target star in the centre and the dark hole surrounding it.

In addition, we combine the dgvAPP360 with an IFS, enabling us to use differential spectrophotometry for high-contrast directly imaged companions. The incoming light is first dispersed into individual spectra, and then recombined into a single `white-light' data point. This has the advantage of smoothing out wavelength-dependent flat-fielding errors and allows wavelength regions with instrumental absorption or highly variable telluric bands to be excluded, meaning systematic effects can be significantly reduced, thus yielding greater stability and precision in the final white-light curve.
\subsection{The HD~1160 system}\label{intro_hd1160}
We test this approach using observations of the HD~1160 system, which is located at a distance of 120.4$\pm$0.6~pc \citep{2016A&A...595A...1G, 2021A&A...649A...1G, 2021AJ....161..147B} and consists of host star HD~1160~A \citep[spectral type A0V;][]{1999mctd.book.....H} and two high-contrast companions, HD~1160~B and C \citep{2012ApJ...750...53N}. HD~1160~B and C lie at separations of $\sim$80~au ($\sim$0.78$\arcsec$) and $\sim$530~au ($\sim$5.1$\arcsec$), respectively. Several key properties of HD~1160~A are listed in Table \ref{table:prim_parameters}. HD~1160~A is bright \citep[K~=~7.040$\pm$0.029 mag,][]{2003yCat.2246....0C}, and the contrast ratio between HD~1160~A and HD~1160~B is $\Delta L^{\prime}=6.35\pm0.12$ mag \citep{2012ApJ...750...53N}. This makes it an ideal target for demonstrating our technique, as a high signal-to-noise detection of the companion allows high cadence monitoring and a deep investigation into any residual systematic effects. A-type stars such as HD~1160~A generally vary below the millimagnitude level, which corresponds to variability amplitudes comfortably below the $\sim$1\% level \citep{2011AJ....141..108C}. We assess the variability of the host star in Section \ref{stellar_variability} using observations from the Transiting Exoplanet Survey Satellite (TESS) mission. The age of the system is poorly constrained, with estimates ranging from 50\raisebox{0.5ex}{\tiny$\substack{+50 \\ -40}$}~Myr \citep{2012ApJ...750...53N} to 100\raisebox{0.5ex}{\tiny$\substack{+200 \\ -70}$}~Myr \citep{2016A&A...587A..56M}. The HD~1160 system may be a member of the Pisces-Eridanus stellar stream, which would place its age at $\sim$120~Myr, but this has yet to be confirmed  \citep{2019AJ....158...77C}.

HD~1160~B has a spectral type close to the brown dwarf/stellar boundary; \citet{2012ApJ...750...53N} found a spectral type of $\sim$L0, but more recent papers suggest that it lies between M5-M7 \citep{2016A&A...587A..56M, 2017ApJ...834..162G, 2020MNRAS.495.4279M}. However, \citet{2020MNRAS.495.4279M} found its spectrum to be highly peculiar and that no spectral model or template in current libraries can produce a satisfactory fit. Although the cause of this peculiarity has not yet been explained, \citet{2020MNRAS.495.4279M} hypothesise that possible causes could include a young system age, dust in the photosphere of HD~1160~B, or ongoing evolutionary processes. The mass of HD~1160~B also remains unclear, primarily due to the poorly constrained age of the system, with estimates ranging from that of a low mass brown dwarf
\citep[$\sim$20~M\textsubscript{Jup},][]{2020MNRAS.495.4279M} to decisively in the stellar mass regime \citep[0.12$\pm$0.01M$_{\odot}\approx123 $M\textsubscript{Jup},][]{2019AJ....158...77C} if the system is indeed a member of the Pisces-Eridanus stellar stream. HD~1160~C is a low-mass star (spectral type M3.5), and its separation of $\sim$530~au ($\sim$5.1$\arcsec$) places it beyond the field of view of most vAPPs currently in use \citep{2016A&A...587A..56M, 2021ApOpt..60D..52D}.

The observations carried out on the HD~1160 system are described in Section \ref{obs}, and in Section \ref{data_red} we describe the spectral extraction and data reduction processes. In Section \ref{phot} we produce and present our differential spectrophotometric light curves of HD~1160~B. We then examine various factors that may be correlated with the light curves and detrend them in Section \ref{lin_reg}. These results and their implications are then discussed in Sections \ref{results} and \ref{discussion}. Lastly, the conclusions of the work are summarised in Section \ref{conclusions}.

\begin{flushleft}
\begin{table}
\caption{Properties of host star HD~1160~A.}
\begin{tabular}{p{0.41\columnwidth}p{0.2\columnwidth}p{0.2\columnwidth}}
\hline
Parameter&Value&Reference(s)\\
\hline
Spectral Type&A0V&(1)\\
Right Acension (J2000)&00:15:57.32&(2)\\
Declination (J2000)&+04:15:03.77&(2)\\
Age (Myr)&10-300&(3, 4)\\
Parallax (mas)&8.2721$\pm$0.0354&(2)\\
Distance (pc)&120.4$\pm$0.6&(2, 5)\\
Proper motion (RA, mas yr$^{-1}$)&20.150$\pm$0.040&(2)\\
Proper motion (Dec, mas yr$^{-1}$)&-14.903$\pm$0.034&(2)\\
Mass (M$_{\odot}$)&$\sim$2.2&(3)\\
T\textsubscript{eff} (K)&9011$\pm$85&(6)\\
log(L/L$_{\odot}$)&1.12$\pm$0.07&(6)\\
log(g) (dex)&$\sim$4.5&(7)\\
{[Fe/H]}&$\sim$solar&(7)\\
V (mag)&7.119$\pm$0.010&(8)\\
G (mag)&7.1248$\pm$0.0004&(2)\\
J (mag)&6.983$\pm$0.020&(9)\\
H (mag)&7.013$\pm$0.023&(9)\\
K (mag)&7.040$\pm$0.029&(9)\\
\hline
\end{tabular}
\textbf{References:} (1) \citet{1999mctd.book.....H}; (2) \citet{2016A&A...595A...1G, 2021A&A...649A...1G}; (3) \citet{2012ApJ...750...53N}; (4) \citet{2016A&A...587A..56M}; (5) \citet{2021AJ....161..147B}; (6) \citet{2017ApJ...834..162G}; (7) \citet{2020MNRAS.495.4279M}; (8) Tycho-2 \citep{2000A&A...355L..27H}; (9) 2MASS \citep{2003yCat.2246....0C, 2006AJ....131.1163S}
\label{table:prim_parameters}
\end{table}
\end{flushleft}
\section{Observations}\label{obs}
We observed the HD~1160 system on the night of 2020 September 25 (03:27:31 - 11:16:14 UT) with the double-grating 360\textdegree{} vector Apodizing Phase Plate (dgvAPP360) coronagraph (see Section \ref{intro_diff_spec}) and the Arizona Lenslets for Exoplanet Spectroscopy (ALES) IFS \citep{2015SPIE.9605E..1DS, 2018SPIE10702E..3LH,  2018SPIE10702E..3FS}. ALES is integrated inside the Large Binocular Telescope Interferometer (LBTI) \citep{2016SPIE.9907E..04H, 2020SPIE11446E..07E}, which works in conjunction with the LBT mid-infrared camera (LMIRcam), on the 2 x 8.4-m LBT in Arizona \citep{2010SPIE.7735E..3HS, 2012SPIE.8446E..4FL}. For these observations, ALES was in single-sided mode and was therefore fed only by the left-side aperture of LBT. Atmospheric turbulence was corrected for by the LBTI adaptive optics (AO) system \citep{2014SPIE.9148E..03B, 2016SPIE.9909E..3VP, 2021arXiv210107091P}. We used the ALES L-band prism, which covers a simultaneous wavelength range of 2.8-4.2~\textmu m with a spectral resolution of R$\sim$40 \citep{2018SPIE10702E..0CS}. The plate scale is $\sim$35~mas~spaxel$^{-1}$. The other LBT aperture was used to feed the Potsdam Echelle Polarimetric and Spectroscopic Instrument (PEPSI), which obtained R~=~50,000 combined optical spectra of HD~1160~A and B in the 383-907nm wavelength range \citep{2015AN....336..324S, 2018SPIE10702E..12S}, which is subject to analysis in other forthcoming works.

Conditions were exceptionally clear and stable throughout the night, with no time lost to weather, and seeing ranged from 0.7-1.4$\arcsec$. We acquired 2210 on-target ALES frames, with an integration time of 5.4~s per frame, giving a total on-target integration time of 11934.0~s ($\sim$3.32~h) spread over $\sim$7.81~h once readout time, nodding, and wavelength calibrations are included. The integration time was chosen such that the stellar PSF remained unsaturated in the core so that it can be used as the photometric reference for the companion. We used an on/off nodding pattern to enable background subtraction, nodding to a position 5$\arcsec$ away for the off-source nod position. As HD~1160~C is located at a similar separation ($\sim$5.1$\arcsec$), we nodded in a direction away from this companion to prevent it from contaminating the frames obtained in the off-source nod position. Beam-switching in this way is possible because of the intrinsic stability provided by the dgvAPP360 coronagraph's  placement in the pupil plane. We obtained dark frames with the same exposure time at the end of the night, and 6 wavelength calibrations were acquired at irregular intervals during the night. LBTI operates in pupil-stabilized mode, such that the field of view was rotating throughout the observations. The total field rotation across the observing sequence was 109.7\textdegree{}. The HD~1160 system was observed from an elevation of 29.4\textdegree{} at the start of the night to a maximum elevation of 61.7\textdegree{}, and then back down to an elevation of 27.5\textdegree{}. The dgvAPP360 creates an annular dark hole around the target PSF, with an inner radius close to the PSF core and an outer radius at the edge of the 2.2$\arcsec$ field of view of the detector (2.7~-~15~$\lambda$/D in ALES mode) \citep{2020PASP..132d5002D, 2021ApOpt..60D..52D}. For these observations, this meant that HD~1160~B was located in the dark hole of HD~1160~A across the entire wavelength range covered by the ALES L-band prism. HD~1160~C (separation~$\sim$5.1$\arcsec$) remained beyond the ALES field of view.

\section{Data reduction} \label{data_red}
The raw ALES images contain the spectra that have been projected onto the detector. Ultimately, we are aiming to produce a light curve for the companion that is made from the `white-light', i.e. combined in the wavelength dimension. To do this, we must first extract the spectra from the raw data along with bad pixel correction and flat-fielding. We can then extract the photometry of the star and the companion at each wavelength before collapsing the data in the wavelength dimension to obtain white-light fluxes. A light curve for the companion is then obtained by dividing the companion flux by the stellar flux to remove systematic trends shared by both. In the following subsections we describe the methods used to carry out each of these steps and obtain the white-light curve of the companion.
\subsection{Spectral data cube extraction}\label{cube_extraction}
Raw ALES data consists of a two-dimensional grid of 63$\times$67 micro-spectra over a 2.2$\arcsec$x 2.2$\arcsec$ field of view, which must be extracted into three-dimensional data cubes of $x$-position, $y$-position, and wavelength~$\lambda$ \citep{2022SPIE12184E..42S}. To do this, we first performed a background subtraction using the sky frames obtained in the off-source nod position. For each ALES image, we subtracted the median combination of the 100 sky frames closest in time. We then extracted the micro-spectra into cubes using optimal extraction, which is an inverse variance and spatial profile weighted extraction approach \citep{1986PASP...98..609H, 2018SPIE10702E..2QB, 2020AJ....160..262S}. The extraction weights were obtained by measuring the spatial profile of each micro-spectrum in the dark-subtracted sky frames. As there is no significant change in the spatial profile as a function of wavelength, we average the spatial profile of each micro-spectrum over wavelength to obtain higher signal-to-noise (S/N). The wavelength calibration of the raw ALES data was then obtained using four narrow-band photometric filters at 2.9, 3.3, 3.5, and 3.9~\textmu m, positioned upstream of the ALES optics. These filters are each of a higher spectral resolution $\frac{\lambda}{\Delta\lambda}$$\sim$100 than ALES, so are therefore unresolved and provide four single-wavelength fiducial spots with which each micro-spectra can be calibrated \citep{2018SPIE10702E..3FS, 2022SPIE12184E..42S}. For each micro-spectrum, we performed this calibration by fitting a second-order polynomial to the calculated pixel positions of these four spots, therefore mapping pixel position to wavelength. Each of the 63$\times$67 micro-spectra was thereby converted into a corresponding spaxel in the three-dimensional data cube \citep{2019AJ....157..244B, 2022AJ....163..217D}. The resulting data cube contained 100 channels in the wavelength dimension ranging from 2.8-4.2~\textmu m.

The primary wavelength calibration used to process this data set was obtained at 08:25:00 UT on the night of observations, i.e. 4 hours 57 minutes into the observing sequence. To test whether a wavelength calibration obtained at a different point in the night has a significant effect on the photometry of the target star and the companion, we also separately processed the data using an alternative wavelength calibration obtained at 04:54:00 UT, 1 hour 26 minutes into the observing sequence. We compare and discuss the results of the two wavelength calibrations in Section \ref{results_wavecal}.

\begin{figure*}
	\includegraphics[scale=0.7]{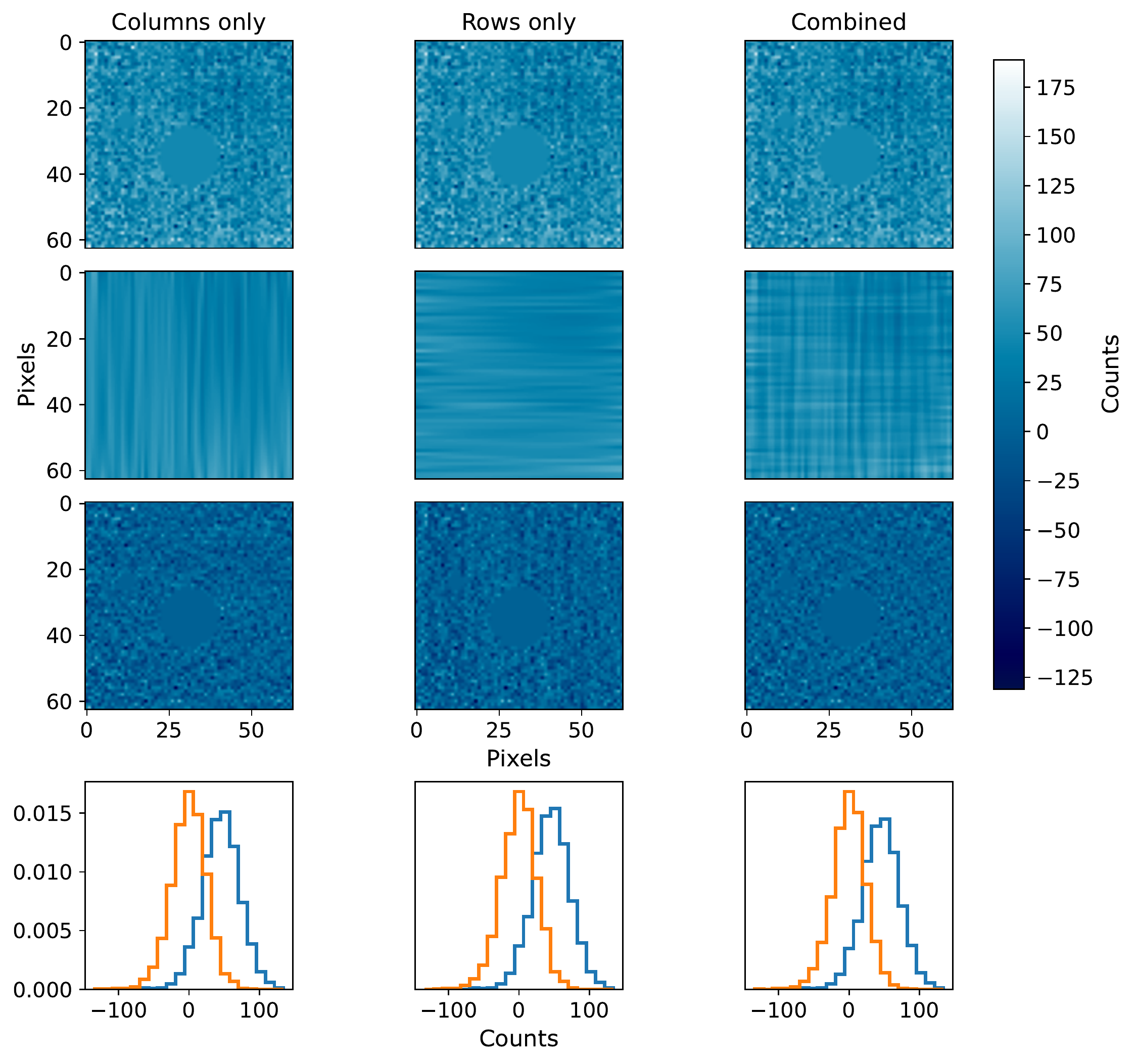}
    \caption{Demonstration of the process for removing systematic row and column discontinuities present in background-subtracted ALES images, as applied to a single frame of data in the 52nd ALES wavelength channel ($\lambda\approx$~3.69~\textmu m). Top panels: input frame prior to the removal of the discontinuities, which are faintly visible as a chequered pattern. All three panels are the same. Both HD~1160~A and B are masked. Second row: results of the third-order polynomial fits individually to the columns (left panel) and rows (centre panel), and the combination of both (right panel). The combination of both was produced by first fitting and removing the column discontinuities, and then repeating the process on the resulting image to fit and remove the rows. We show row and column fits to the input frame separately here to highlight their individual contribution to the original systematics. The star and the companion were masked during this process, and the values to be removed at their locations were found through interpolation of the fits. Third row: data frame with the discontinuities removed by subtracting the fits. The histograms in the bottom row show the distributions of the counts in the unmasked regions of the third row images, with the original noise distribution in blue and the noise distribution after the discontinuities were removed in orange. The bottom right panel shows the version used in the analysis.}
    \label{fig:discont}
\end{figure*}

\subsection{Data processing}\label{data_processing}
Once the spectra were extracted into background-subtracted three-dimensional cubes of images for each exposure, we applied several data reduction steps to remove systematics and improve the S/N at the location of the companion. We first removed 8 time frames from the data in which the AO loop opened while the data was being collected. Next, we identified bad pixels using a 6$\sigma$ filter and replaced them with the mean of the neighbouring pixels. We then applied a flat-field correction to calibrate the data against the response of the detector. For each wavelength channel, the corresponding flat was a time-average of the frames obtained in the `off' nod position, which had then been corrected for bad pixels in the same way as the science frames and smoothed over using a Gaussian filter. These flats were then divided by the maximum value in the frame such that the value of every pixel was between zero and one. We then divided the science frames by these smoothed, normalised sky flats. This flat-fielding process was also repeated separately using a median filter instead of a Gaussian filter as a means to test the robustness of this step in our method. We proceed with the Gaussian filter and discuss the impact of the choice of flat frame on the photometry of the star and companion in Section \ref{results_wavecal}.

\citet{2022AJ....163..217D} previously identified that background-subtracted, flat-fielded ALES images contain residual structure that cannot be described by purely Gaussian noise, in the form of time-varying row and column discontinuities (faintly visible in the top panel of Figure~\ref{fig:discont}). Such discontinuities are expected and arise from the way in which the micro-spectra lie across multiple channels of the LMIRcam detector \citep{2022AJ....163..217D}. We followed the method of \citet{2022AJ....163..217D} to characterise and remove these discontinuities by fitting a third-order polynomial to each row and column in each frame (Figure~\ref{fig:discont}). Removing these systematics is important as they could impact the precision of our differential light curve, or even generate a false variability signal if the target moves over them throughout the observing sequence. Prior to fitting, we applied circular masks at the locations of the star and the companion in each frame such that their flux did not contaminate the fit. To find the position of the star in each time frame, we selected a wavelength channel with a high stellar flux per frame (channel 52, $\lambda\approx$~3.69~\textmu m) and fit the PSF core with a 2D Gaussian. The position of the companion in each time frame was then identified using its separation and position angle relative to the star and accounting for the effect of the field rotation over time. We then masked the star and the companion across all wavelength channels using circular masks with diameters of 18 pixels and 5 pixels, respectively, before fitting the third-order polynomials to each column. The resulting values were then subtracted from the data to remove the column discontinuities. This process was then repeated for each row in the resulting image to remove the row discontinuities.

In addition to removing these systematic discontinuities, this process has the effect of removing residual background flux not eliminated by earlier processing steps. This is indicated by the histograms in Figure~\ref{fig:discont}, which show that the noise distribution of the data was offset from zero prior to the removal of the discontinuities (in blue) but is approximately consistent with zero after this process has been applied (in orange). We then used the position of the star in each frame, found when applying the masks in the previous step, to spatially align the data such that the star was in the centre of each frame. Finally, we rotationally aligned the images by applying an anticlockwise rotation corresponding to their parallactic angles.

We did not use any further post-processing methods that reduce quasistatic speckle noise through the subtraction of reference PSFs, such as Angular Differential Imaging \citep[ADI,][]{2006ApJ...641..556M}. Although the field rotation over the night of observations was sufficient enough to use these to remove noise with minimal companion self-subtraction, doing so would also remove the unsaturated stellar reference PSF, which is required to eliminate systematics in the companion photometry. It would not be possible to use the host star PSF prior to ADI subtraction as a photometric reference for the companion PSF after ADI subtraction, as the two would no longer share the same systematic trends. Furthermore, HD~1160~B is sufficiently bright that it can be detected at ample S/N for our purposes without further noise reduction.

\begin{figure*}
\centering
  \textcolor{white}{\frame{\includegraphics[scale=0.58]{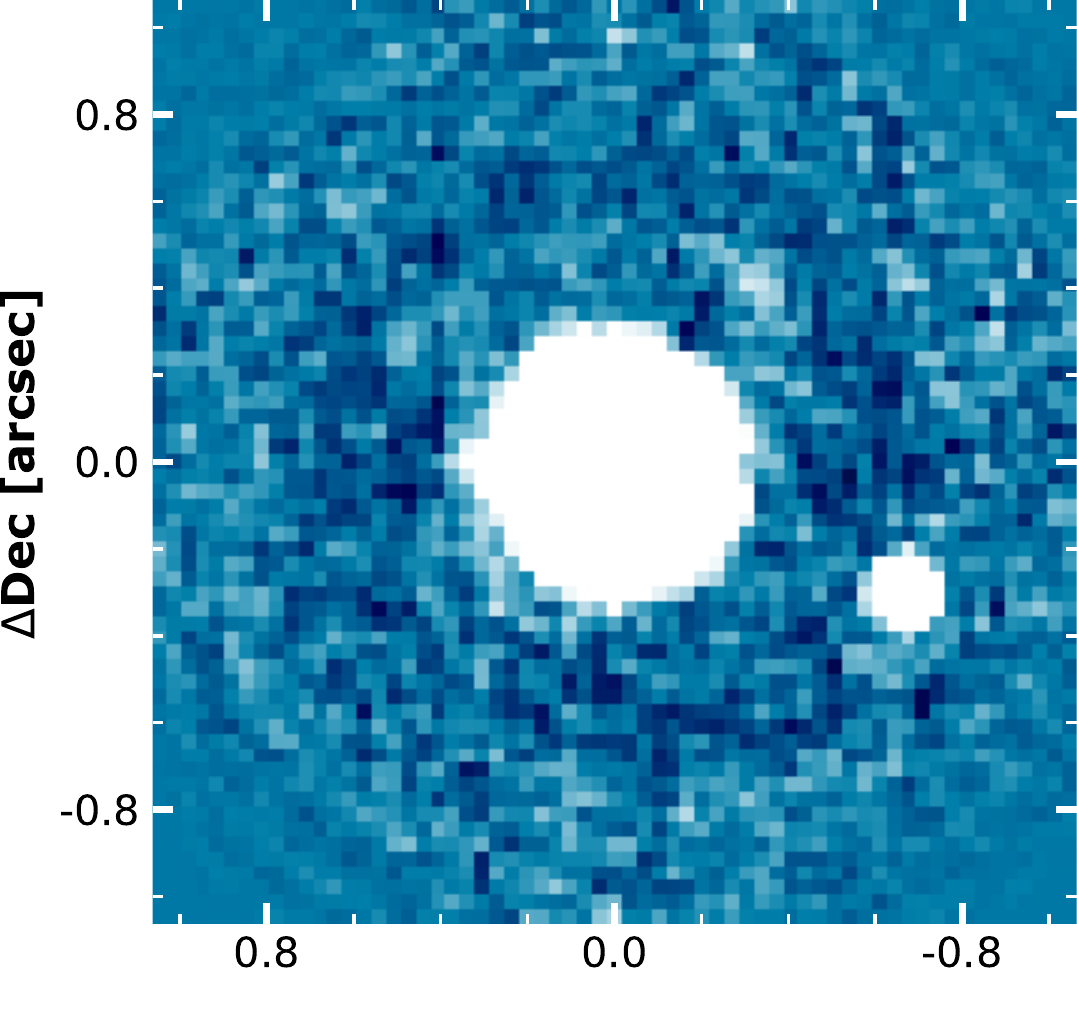}}}%
  \textcolor{white}{\frame{\includegraphics[scale=0.58]{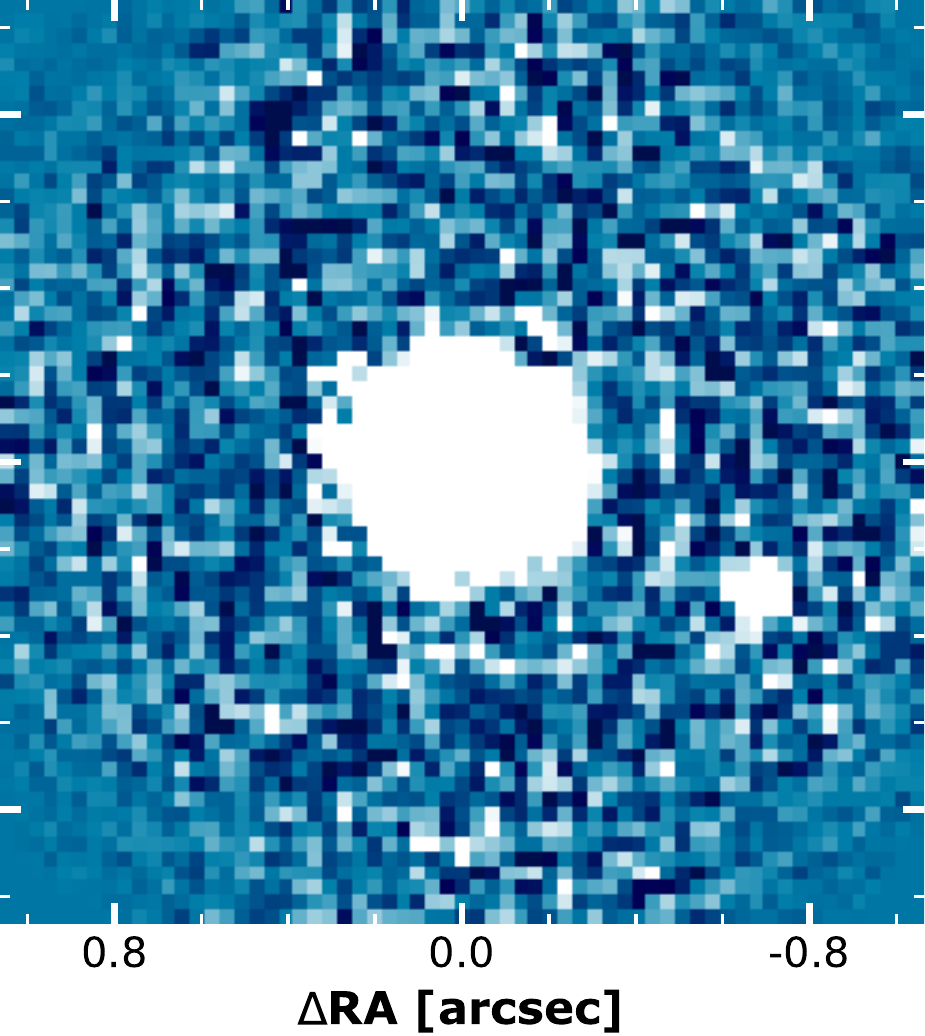}}}%
  \textcolor{white}{\frame{\includegraphics[scale=0.58]{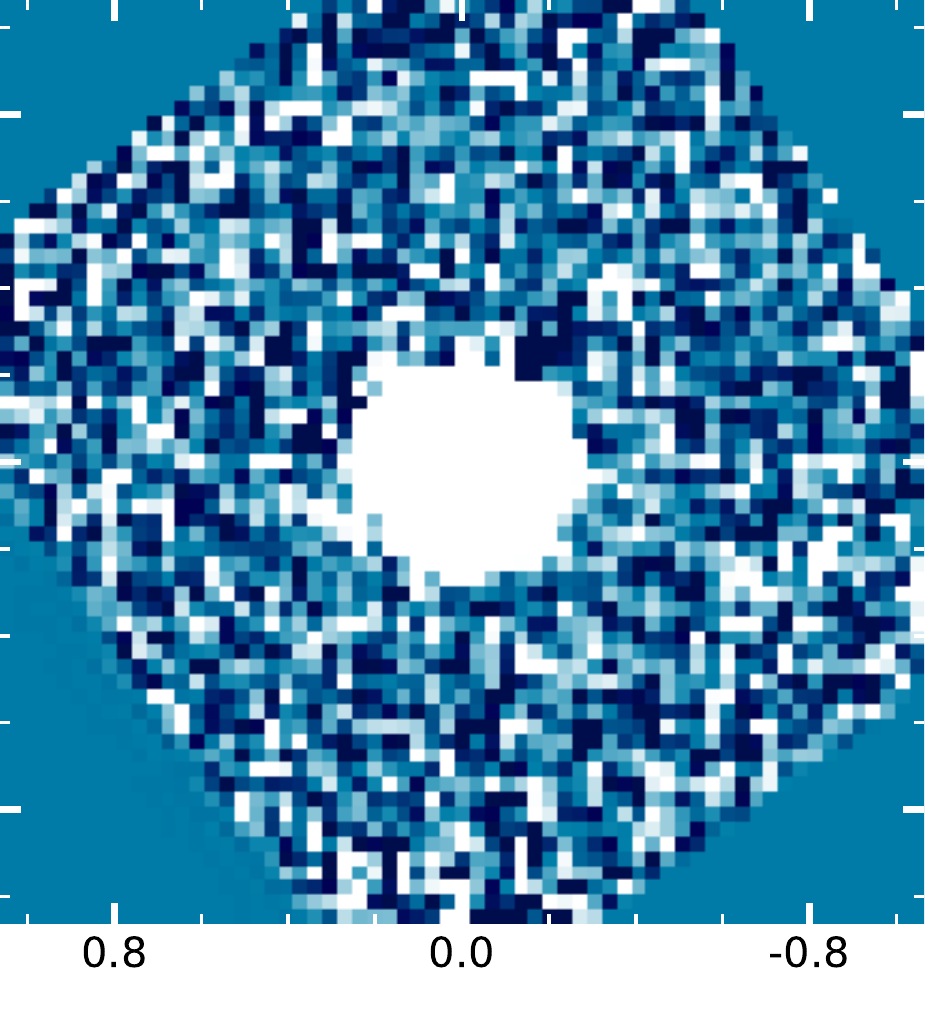}}}%
\caption{Reduced images of the HD~1160 system obtained with LBT/ALES and the dgvAPP360 coronagraph after all data processing steps and wavelength channel selection. Left: the median combination, in both wavelength and time, of all wavelength channels in the 3.59-3.99~\textmu m range, covering a total integration time of 11891~s ($\sim$3.31~h). Centre: example image from the same data cube but median combined only in the time dimension, along the 3.64~\textmu m wavelength channel. Right: example time frame (integration time = 5.4 seconds) resulting from a median combination in wavelength over the 3.59-3.99~\textmu m wavelength range. All three images use the same arbitrary logarithmic colour scale, and are aligned to north. North is up, and east is to the left.}
\label{fig:final_images}
\end{figure*}

\subsection{Wavelength channel selection}\label{channel_selection}
Although the final processed cubes contain data from 100 wavelength channels across the observed wavelength range of 2.8-4.2~\textmu m, not all of these channels are suitable for further analysis. The first 3 and final 10 channels contain flux from the adjacent spaxel in the dispersion direction as an oversized spectral length is required to extract the spectrum at each position, so the extracted data overlaps slightly in the wavelength dimension \citep{2022SPIE12184E..42S}. Furthermore, the dgvAPP360 contains a glue layer that causes up to 100\% absorption between 3.15 and 3.55~\textmu m \citep{2017ApJ...834..175O, 2021ApOpt..60D..52D}. As described in Section \ref{intro_diff_spec}, one of the key advantages of a spectrophotometric approach is the option to exclude channels that are known to cause systematic variability in the `white-light' curve, hence improving the companion S/N and stability in the combined image when compared to combining all wavelength channels without any selection. This is key to reducing large systematic effects that may otherwise dominate the variability signals that we are aiming to measure. For the purposes of this technique demonstration, we proceed using 30 sequential wavelength channels (45-74, spanning 3.59-3.99~\textmu m) which all have a high throughput and do not lie in the regions affected by the dgvAPP360 glue absorption, significant telluric absorption, or the overlapping spectral traces. In the left panel of Figure \ref{fig:final_images} we show the median combination of these channels in both wavelength and time, while the centre and right panels respectively show example images from the median combined cubes in the time and wavelength dimensions only. The images shown are those processed using the flat frame that was smoothed using a Gaussian filter; the equivalent images as processed using the median-smoothed flat frame are visually indistinguishable from these.
\section{Generating differential spectrophotometric light curves}\label{phot}
Variability arising from instrumental systematics and the effects of Earth's atmosphere, such as airmass, seeing, and tellurics, contaminate the raw flux of the companion. Simultaneous flux measurements of a photometric reference are required to eliminate this contaminant variability and produce a differential light curve of the companion, relative to the photometric reference. Although suitable photometric references are generally absent when using coronagraphs (Section \ref{intro}), the dgvAPP360 coronagraph uniquely provides an image of the host star simultaneously to the companion, allowing the star to be used as the photometric reference when it is not saturated. Its placement in the pupil plane also makes it inherently stable and insensitive to tip/tilt instabilities \citep{2017ApJ...834..175O, 2022AJ....163..217D}.
\subsection{Aperture photometry}\label{aper_phot}
\begin{figure}
	\includegraphics[scale=0.77]{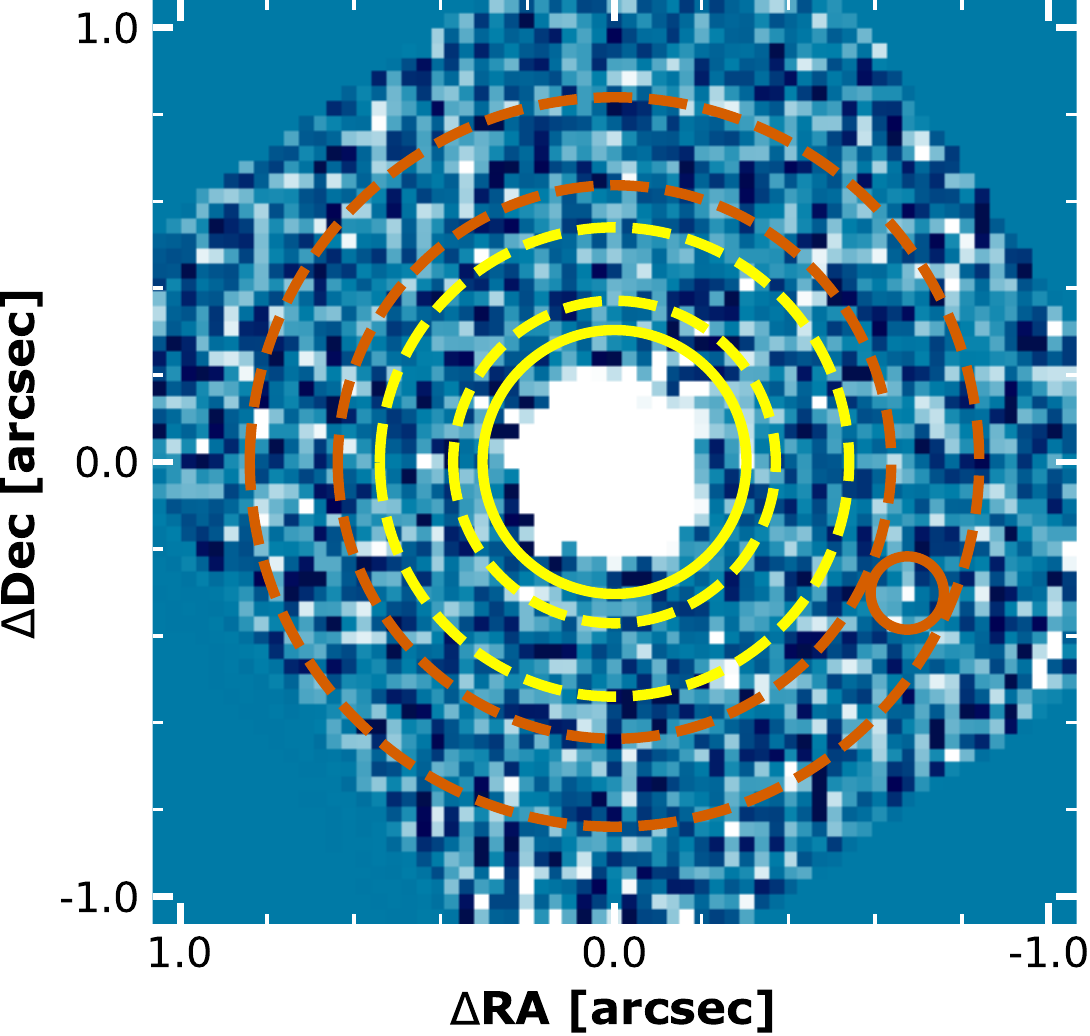}
    \caption{The apertures (continuous lines) and annuli (dashed lines) used to extract photometry and background measurements for host star HD~1160~A (in yellow) and companion HD~1160~B (in orange). The image is a single time frame from the 52nd ALES wavelength channel ($\lambda\approx$~3.69~\textmu m). North is up, and east is left. The orange aperture is placed at the location of HD~1160~B, which is too faint to be visible in a single frame. The companion was masked when extracting photometry in the annulus for the companion background.}
    \label{fig:apertures_plot}
\end{figure}

\begin{figure*}
	\includegraphics[scale=0.75]{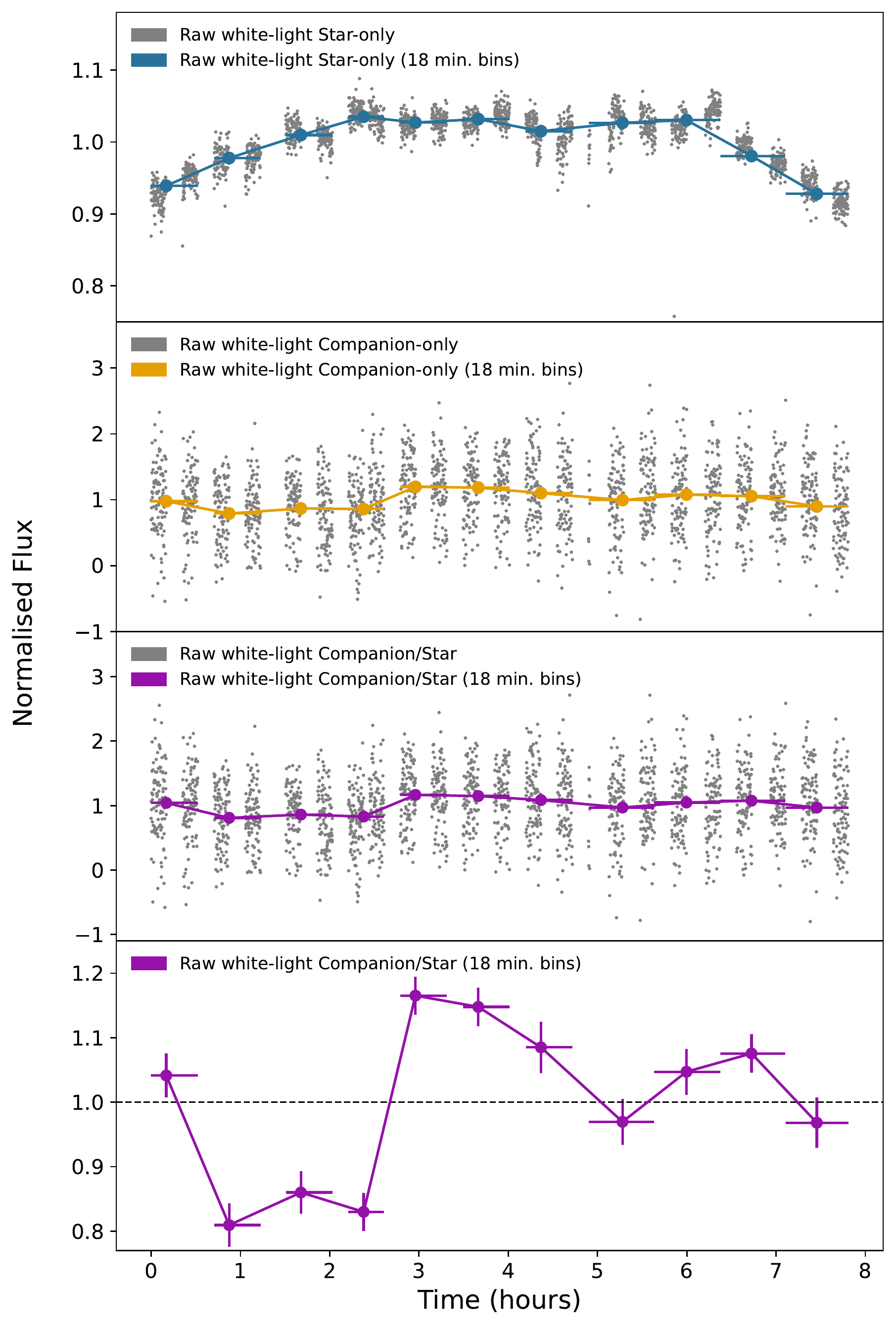}
    \caption{The top two panels show the normalised raw white-light flux ($\lambda=$3.59-3.99~\textmu m wavelength range) of host star HD~1160~A (in grey, top panel) and companion HD~1160~B (in grey, second panel). The blue and orange lines are the same fluxes of the star and companion, respectively, binned to 18 minutes of integration time per bin. Each has been normalised by dividing by the mean value across the full sequence. The third panel shows, in grey, the resulting differential white-light curve when the white-light of the companion is divided by the white-light of the star to remove trends shared by both. The same data binned to 18 minutes of integration time per bin is shown in purple. The light curve in the bottom panel is the same as the third panel but zoomed in on the y-axis with a dashed line at normalised flux~=~1 for clarity. Provided that there is no contamination from stellar variability, variations in this light curve are a combination of any intrinsic companion variability and trends arising from non-shared systematics. The gaps in the data are due to the two-point on/off nodding pattern used to collect sky frames for background subtraction.}
    \label{fig:raw_lcs}
\end{figure*}

We used version 1.4.0 of the Photutils Python package \citep{larry_bradley_2022_6385735} to simultaneously extract aperture photometry of HD~1160~A and B. We carried out this process for every individual frame in each of the 30 wavelength channels in the 3.59-3.99~\textmu m range chosen in Section \ref{channel_selection}, with the aim of then combining these in the wavelength dimension to produce the white-light flux for each object. Circular apertures with radii of 9 pixels (3.1~$\lambda$/D) and 2.5 pixels (0.9~$\lambda$/D) were used for the host star and the companion, respectively. To estimate the background flux at the position of the star, we also extracted photometry in a circular annulus centred on the stellar location with inner and outer radii of 11 and 16 pixels, respectively. It was not possible to use this method to estimate the background flux at the position of the companion as the companion lies close to the edge of the field of view in some frames, limiting the space available to place an annulus that would be statistically wide enough. We therefore instead followed an approach used by \citet{2021MNRAS.503..743B}, estimating the background at the companion location by masking the companion and extracting flux in a circular annulus centred on the host star, with a width of 6 pixels at the radial separation of the companion. As most of the residual background in each frame was eliminated by the data processing steps in Section \ref{data_processing}, these background values are close to zero. These apertures and annuli are shown in Figure \ref{fig:apertures_plot}, overlaid on a single processed time frame of data in the 52nd ALES wavelength channel ($\lambda\approx$~3.69~\textmu m). We then removed the residual background from our stellar and companion flux measurements by multiplying the mean flux per pixel in the background annuli by the area of the corresponding apertures and subtracting the resulting values from the aperture photometry. We then produced single white-light measurements for both the companion and the star at each time frame by taking the median combination of the photometric measurements across the 30 wavelength channels. These raw time series, uncorrected for shared variations introduced by Earth's atmosphere (i.e. before division), are shown in grey in the top two panels of Figure \ref{fig:raw_lcs}. The discrete gaps in integration reflect time spent off-target due to the two-point on/off nodding pattern used to enable background subtraction. We also plot the data binned to 18 minutes of integration time. We binned the data by taking the median value in each time bin. The error on the binned fluxes are the Gaussian approximation of the root mean square (RMS) i.e. median absolute deviation (MAD) $\times$ 1.48 of the flux measurements inside each bin divided by $\sqrt{N-1}$, where $N$ is the number of frames per bin. Next, we removed variability due to Earth’s atmosphere and other systematics from the unbinned raw flux of the companion using the unbinned raw flux of the host star, which acts as a simultaneous photometric reference. By dividing the unbinned companion flux by the unbinned stellar flux, we eliminate trends common to both and produce a differential light curve that only contains non-shared variations. Assuming that the host star is not itself varying (see Section \ref{stellar_variability}), the resulting differential light curve reflects the intrinsic variability of the companion plus any contamination arising from non-shared systematics. We show this raw differential light curve in the third panel of Figure \ref{fig:raw_lcs}. The bottom panel shows a zoomed-in view of the binned version, with tighter limits on the y-axis.

In the following sections we examine a number of physical, instrumental, and processing factors that may be correlated with non-astrophysical features in the differential white-light curve, and in Section \ref{lin_reg} we model and remove non-shared variations arising from some of these factors.

\subsection{TESS light curves of host star HD~1160~A}\label{stellar_variability}
Although the vast majority of A-type stars generally vary well below the $\sim$1\% level, a small fraction vary at a much higher level, showing up to $\sim$15\% variations \citep{2011AJ....141..108C}. To test our assumption that the host star HD~1160~A is not varying at a level that impacts our differential light curve, we used data from the TESS mission, which is publicly available on the Mikulski Archive for Space Telescopes (MAST) data archive\footnote{MAST data archive portal: \url{https://mast.stsci.edu/portal/Mashup/Clients/Mast/Portal.html}}. The TESS mission observed HD~1160~A for 25 days in Sector 42 (from 2021 August 21 to 2021 September 14) and for 26 days Sector 43 (from 2021 September 16 to 2021 October 11), 51 days in total, with 2~minute cadence. The TESS detector bandpass covers a broad-band wavelength range of 0.6-1.0~\textmu m \citep{2015JATIS...1a4003R}, which does not overlap with our LBT/ALES observations in the 2.8-4.2~\textmu m range. However, as stars are generally less variable in the infrared than in the optical regime, any variations in the TESS light curve of HD~1160~A should represent an upper limit for its variability at the wavelengths covered by ALES \citep[e.g.][]{1998A&A...329..747S, 1999A&A...345..635U, 2004A&ARv..12..273F, 2012ApJ...748...58D, 2012MNRAS.427.3358G, 2013ACP....13.3945E, 2022arXiv220109905R}. Each TESS pixel covers 21$\arcsec$ on sky. This means that HD~1160~A, HD~1160~B (at a separation of $\sim$0.78$\arcsec$), and HD~1160~C (at a separation of $\sim$5.1$\arcsec$) are not resolved separately in the TESS images and appear as a single object. However, as both HD~1160~B ($\Delta J=8.85\pm0.10$ mag, $\Delta L^{\prime}=6.35\pm0.12$ mag) and C ($\Delta J=6.33\pm0.04$ mag, $\Delta L^{\prime}=4.803\pm0.005$ mag) are far fainter than HD~1160~A, especially at shorter wavelengths, it is reasonable to assume that flux of HD~1160~A will dominate in the TESS data \citep{2012ApJ...750...53N}.

\begin{figure}
	\includegraphics[width=\columnwidth]{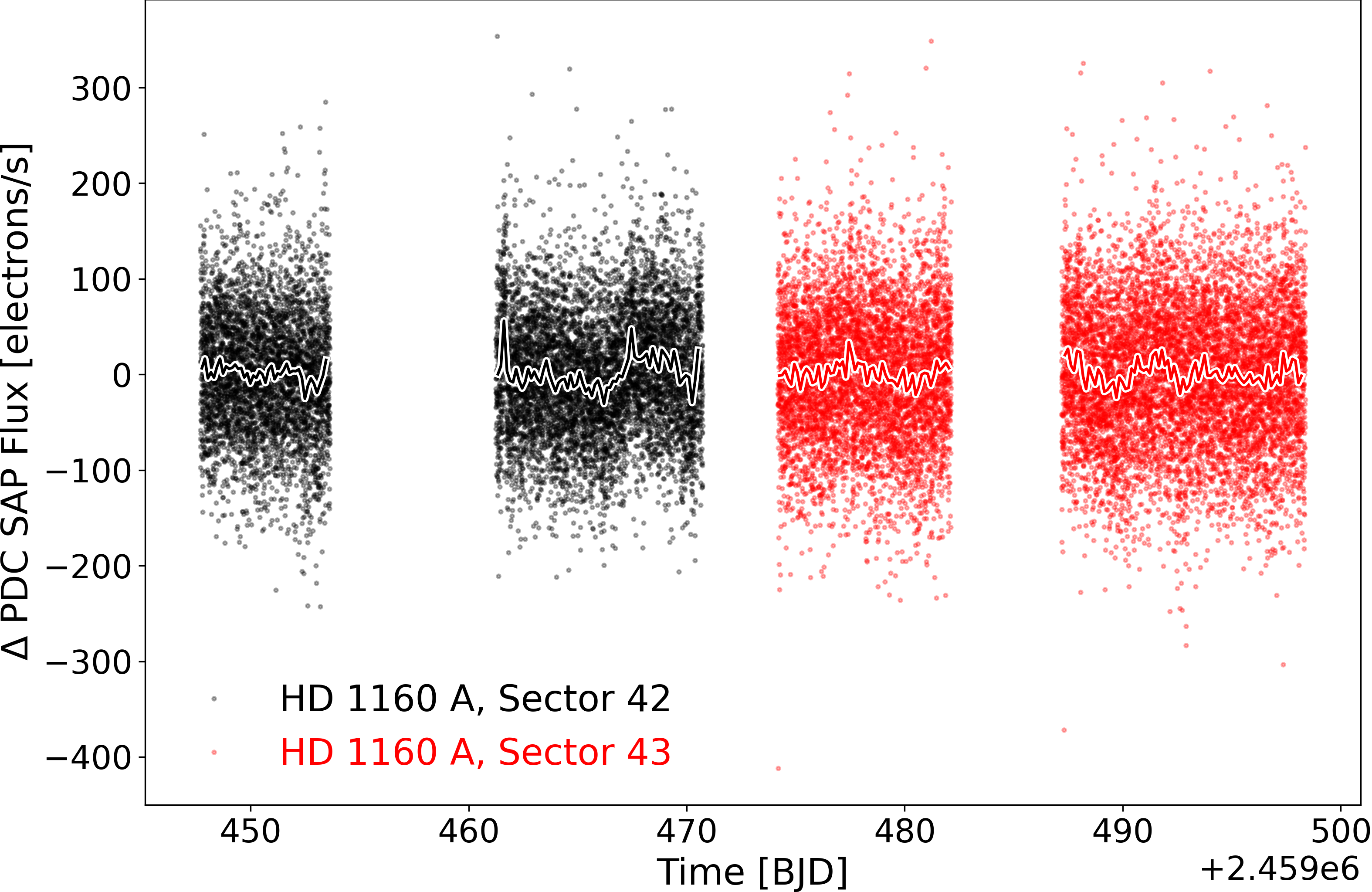}
    \caption{TESS 2 minute cadence PDC SAP light curve of HD~1160~A obtained from TESS sectors 42 (black points) and 43 (red points). Light curves from both sectors have been normalized to the median count level for the respective sector to remove the systematic change in the base flux level from one sector to the next. Overplotted is the binned TESS light curve for visual aid. The standard deviation of the stellar flux is 0.027\% and 0.03\% for sectors 42 and 43, respectively.}
    \label{fig:tess_plot}
\end{figure}

We first masked out any bad quality exposures using the one-hot encoded quality mask in the `QUALITY' keyword in the header of the light curve files provided by the TESS Science Processing Operations Center \citep[SPOC,][]{Jenkins2016} on MAST. We then used the `CROWDSAP' keyword in the header to get an estimate of the ratio of target flux to total flux in the optimal aperture used for the PDC SAP (Pre-search Data Conditioning Simple Aperture Photometry) flux \citep[e.g.][]{2022MNRAS.515.5018P}. The `CROWDSAP' value for sector 42 and 43 indicates that 0.17\% and 0.2\% flux is from dilution by nearby sources. We subtract the estimated diluted flux from each exposure in both the sectors. The resultant light curves for both the sectors are shown in Figure \ref{fig:tess_plot}. Although the TESS observations are not contemporaneous with our LBT/ALES observations, we do not see variations above 0.03\% in the light curve of HD~1160~A over the timescale covered by the two TESS sectors (51 days). As this is far smaller than the precision of our differential light curve, we proceed with the assumption the host star HD~1160~A is non-varying within the flux precision of our analysis of the variations in the light curve.

\subsection{Impact of wavelength calibration and flat-field smoothing}\label{results_wavecal}
\begin{figure*}
	\includegraphics[scale=0.77]{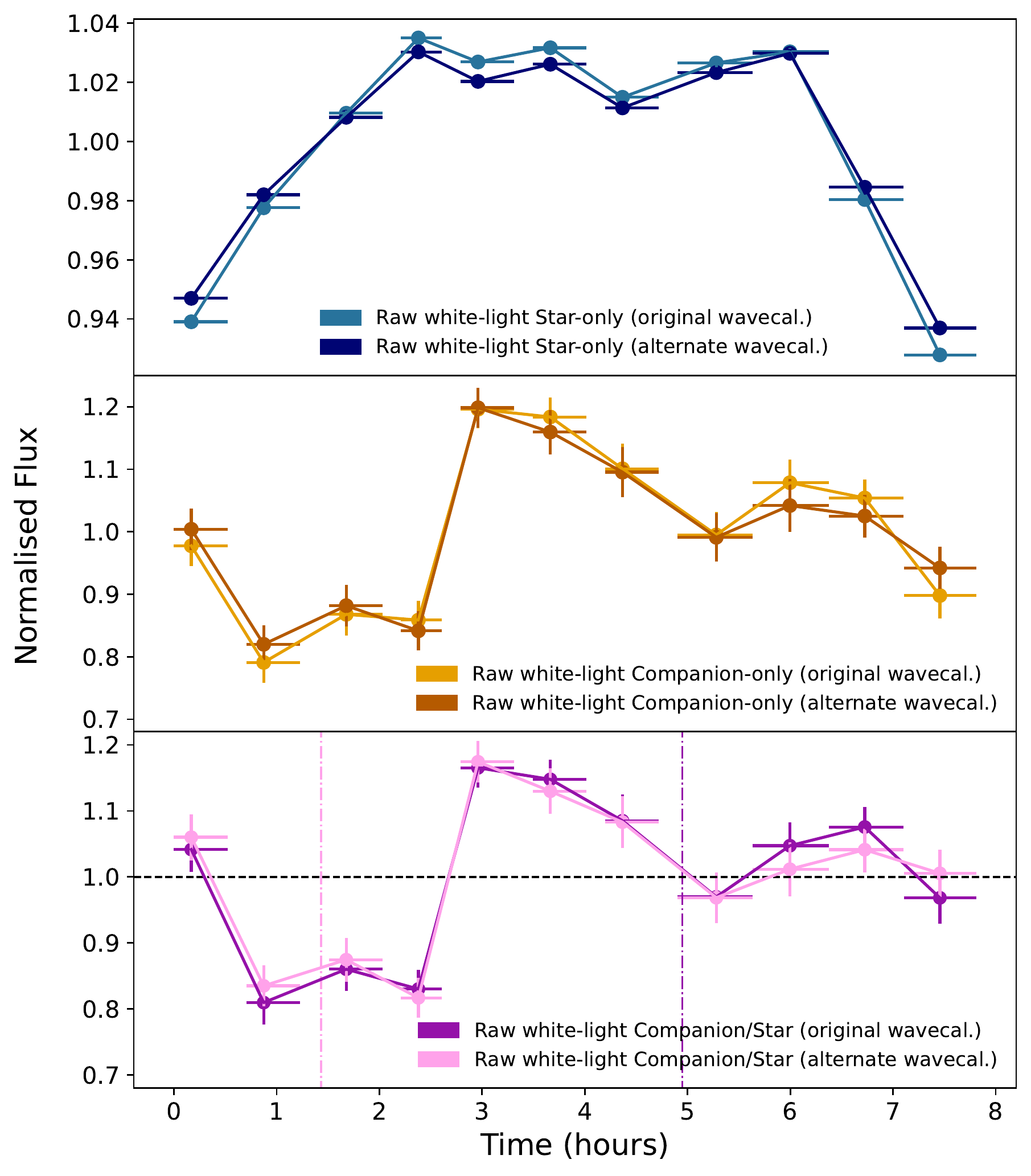}
    \caption{Impact of different wavelength calibrations on the star-only and companion-only fluxes, and the differential white-light curve, shown in 18 minute binning. The alternate wavelength calibration was chosen as the one that diverged most from the one originally used. The times at which the wavelength calibrations were obtained are indicated by vertical dashed lines in the same colours as the corresponding light curve.}
    \label{fig:lc_wcal_comparison}
\end{figure*}

In Section \ref{cube_extraction}, we described the process used to perform the wavelength calibration of the raw data and to extract the micro-spectra into a three-dimensional image cube. This step was repeated separately using the wavelength calibration that was the most divergent of the 6 obtained throughout the observing sequence, i.e. the 3.9~\textmu m fiducial spots for this wavelength calibration were the most significantly offset compared to the one that was originally used. Over the course of the night, the projection of the micro-spectra onto the detector drifts slightly. If this drift is significant then a particular wavelength calibration may not remain accurate for the entire observing sequence, potentially producing a false variability signal when the wrong part of the spectrum is assigned to a given channel. Repeating our spectral extraction using a wavelength calibration obtained at a different point during the observations allows us to test whether this effect has a significant impact on the photometry of the target star and the companion. After extracting the micro-spectra using the alternative wavelength calibration, we then processed the data again in full to produce an alternative differential light curve. The original and alternative wavelength calibrations were obtained at 4 hours 57 minutes (08:25:00 UT) and 1 hour 26 minutes (04:54:00 UT) after the beginning of the observing sequence, respectively. We plot the resulting alternative stellar and companion fluxes, and the differential white-light curve (binned to 18 minutes) in Figure \ref{fig:lc_wcal_comparison}, alongside the originals from Figure \ref{fig:raw_lcs} for comparison. The differential light curves in each case are consistent within 1$\sigma$, indicating that the extracted photometry is sufficiently robust to changes in the wavelength calibration and sub-pixel mis-registration of the spatial profiles for each microspectrum.

\begin{figure}
	\includegraphics[scale=0.47]{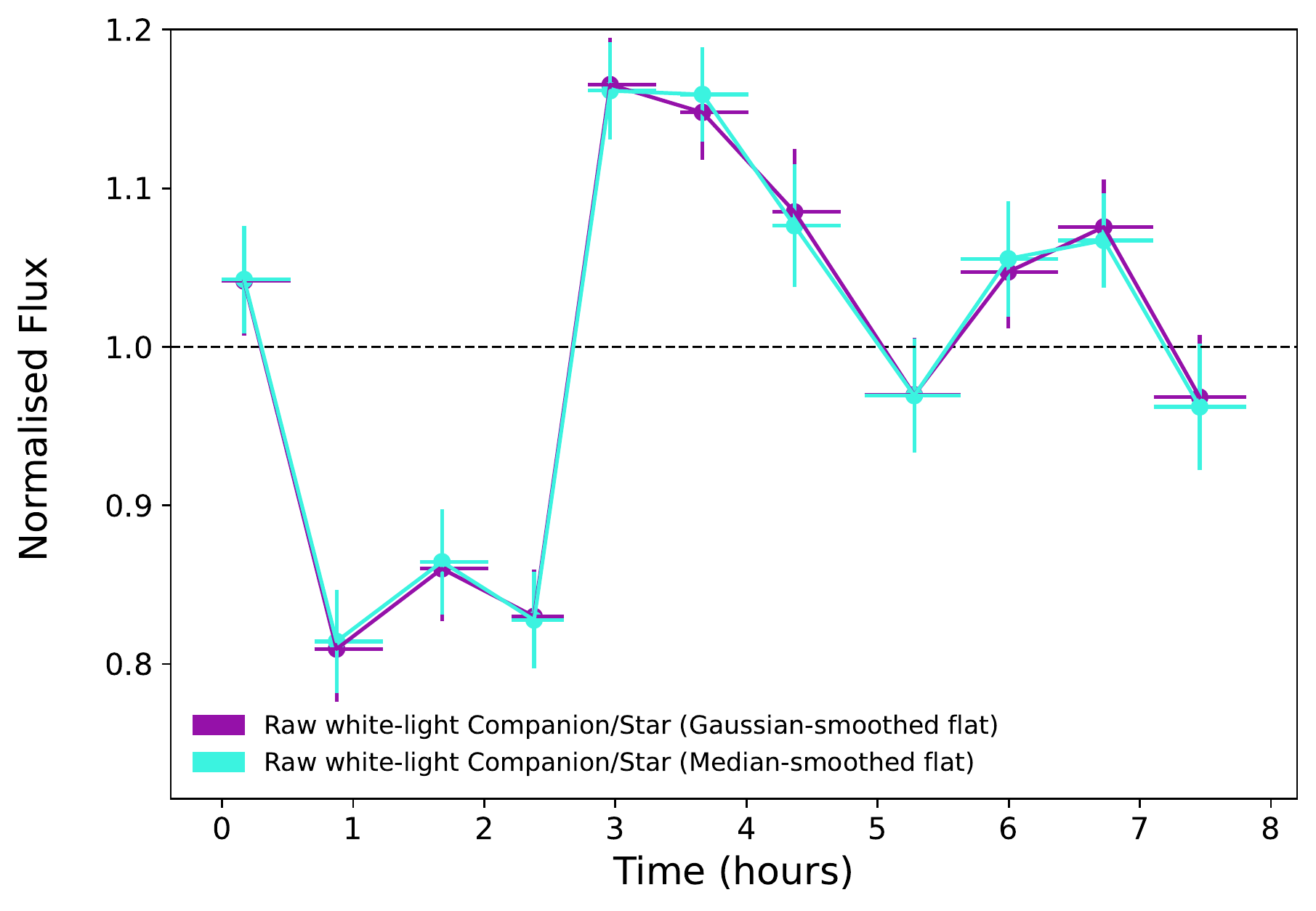}
    \caption{The raw differential companion/star white-light curves that are produced when the flat-field correction uses a flat that was smoothed using a median filter and a Gaussian filter, in turquoise and purple, respectively. The latter light curve is the same as that in the bottom panel of Figure \ref{fig:raw_lcs}, reproduced here for comparison.}
    \label{fig:lc_flat_choice}
\end{figure}

In Section \ref{data_processing}, we described our method for applying a flat-field correction to the data to calibrate for the non-uniform response of the detector. Incorrect flat-fielding can lead to a false variability signal if the companion moves over regions of the detector with a non-uniform response that has not been properly calibrated. To test the robustness of our differential light curve to differences in the flat used we processed the data in two separate streams, using a flat that had been smoothed over using a Gaussian filter and a median filter, respectively. We plot the resulting differential light curves in Figure \ref{fig:lc_flat_choice} for comparison. The two differential light curves are in close agreement and every binned data point lies well within their 1$\sigma$ error bars, indicating that the method for producing the flat is robust and does not significantly affect the final images or extracted photometry of the star or companion.

\section{Detrending through linear regression}\label{lin_reg}
Trends shared by the star and companion fluxes are removed in the differential light curve (see bottom panel of Figure \ref{fig:raw_lcs}), but any non-shared trends will still be present, including the intrinsic companion variability signal that we aim to measure. To improve our sensitivity to the companion's variability, we needed to remove the non-astrophysical residual trends in the differential light curve, which can arise from both telluric and instrumental sources. In the field of exoplanet transmission spectroscopy, such systematic trends are generally modelled and removed from light curves using either a polynomial model created by simultaneously fitting several decorrelation parameters \citep[e.g.][]{2009A&A...493L..35D, 2011A&A...528A..49D, 2012A&A...545L...5B, 2014AJ....147..161S, 2018AJ....156...42D, 2020AJ....160...27D, 2019A&A...631A.169T}, or a non-parametric model produced using Gaussian processes \citep[e.g.][]{2012MNRAS.419.2683G, 2013MNRAS.428.3680G, 2013ApJ...772L..16E, 2015MNRAS.451..680E, 2016ApJ...822L...6M, 2020MNRAS.494.5449C, 2020AJ....160..188D, 2022MNRAS.510.3236P, 2022MNRAS.515.5018P}. Unlike traditional transmission spectroscopy observations, our target is significantly fainter than the simultaneous reference that we use for detrending. Furthermore, the target was not pixel-stabilised for these observations and moved across the detector throughout the night, so we might predict that the measured light curves could be significantly correlated with the change in position of the companion and the star on the detector over time. Knowing how to remove these systematics is key to obtaining high-precision light curves in future observations of directly imaged exoplanets. For space-based observations the instrumentation is generally sufficiently stable such that the systematics are repeatable over time, allowing them to be well characterised. This is more challenging for ground-based observations, like those here, which are inherently less stable as Earth's atmosphere introduces systematics that can vary night by night.

As a basic demonstration of how to remove such residual trends from ground-based differential light curves of directly imaged planets, we here used a multiple linear regression approach to simultaneously fit several possible sources of systematics. This is not intended as a strictly rigorous statistical analysis of the trends in the light curve, but is done to perform an initial investigation into which parameters have the greatest impact and to illustrate an example approach of how to do this for future observations. As studies move towards increasing precision to measure smaller amplitude variability, one might instead consider approaches using Gaussian processes or similar.

\begin{figure*}
	\includegraphics[scale=0.53]{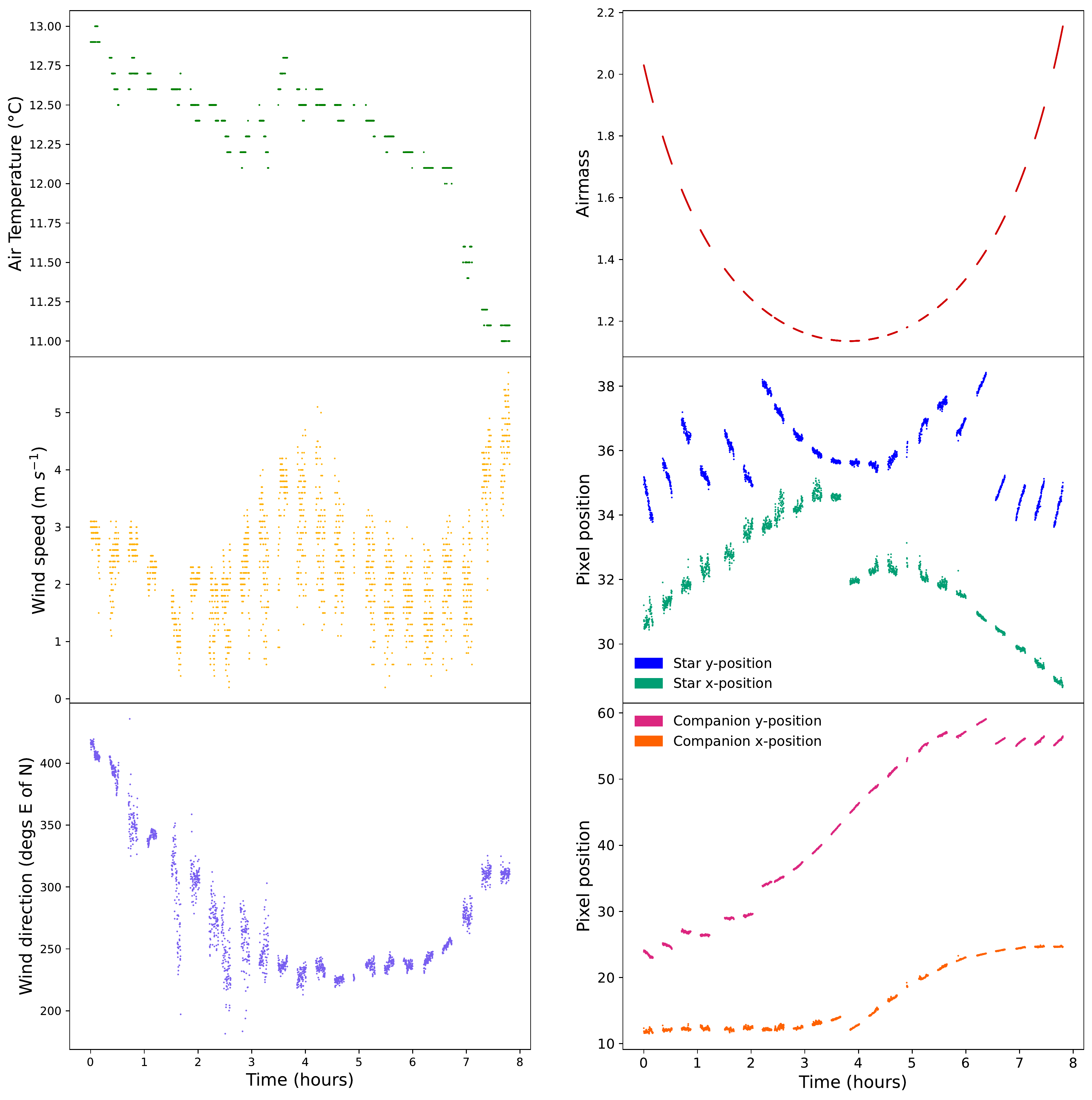}
    \caption{The decorrelation parameters used when detrending the differential light curve via linear regression. From top to bottom, the left panels show the air temperature (in \textcelsius), wind speed (in m~s$^{-1}$), and wind direction (in degrees east of north) at the observatory as a function of time, as extracted from the FITS headers of the raw data. The panels on the right show airmass, the x- and y-positions of the star in pixels, and the x- and y-positions of the companion in pixels, as a function of time. The gaps in time reflect the on/off nodding pattern of the observations. The companion and star positions are those in the images prior to spatial and rotational alignment, and the sharp discontinuities in pixel position are due to manual offsets applied to keep the star close to the centre of the small field of view. The stellar positions follow arc-shaped trends aside from these discontinuities, which correlate with the pointing altitude of the telescope and arise from flexure of the ALES lenslet array as the telescope rotates throughout the night. The change in the companion position has an additional trend due to the 109.7\textdegree{} rotation of the field of view over the observing sequence.}
    \label{fig:lin_reg_params}
\end{figure*}

An investigation of the LBT telemetry and white light images revealed eight physical and instrumental factors, shown plotted against time in Figure~\ref{fig:lin_reg_params}, that varied notably during the observing sequence and may be correlated with the residual trends in the differential light curve. We therefore included these parameters in the linear regression. The first three of these were air temperature, wind speed, and wind direction, shown in the three panels on the left-hand side of Figure~\ref{fig:lin_reg_params}. We also considered airmass, which is shown in the top-right panel. While the light from the companion and its host star pass through almost identical airmass (maximum difference $\sim$10$^{-5}$), their significantly different colours mean that atmospheric extinction due to absorption and e.g. Rayleigh scattering can result in a differing airmass dependence, even when such scattering effects are reduced at our longer 2.8-4.2~\textmu m wavelength range \citep{1955asqu.book.....A, 2005AN....326..134B, 2015ApJ...802...28C, 2022MNRAS.510.3236P}.

The remaining four parameters included in the linear regression were the x- and y-  pixel positions of the star and the companion in the images prior to spatial and rotational alignment, shown in the centre-right and bottom-right panels of Figure~\ref{fig:lin_reg_params}. The dgvAPP360 is located in the pupil plane, meaning that drifts in the locations of the target PSFs on the detector do not affect its response and performance as it applies the phase modification to every source in the field of view \citep{2017ApJ...834..175O, 2022AJ....163..217D}. However, systematics could still be introduced by such drifts if there are variations in the instrumentation or detector response. A number of sharp discontinuities in the y-position, and a singular discontinuity in the x-direction, can be seen in Figure \ref{fig:lin_reg_params}. These discontinuities are the result of manual positional offsets applied during the observations to keep the star close to the centre of the small ($\sim$2.2$\arcsec$x 2.2$\arcsec$) ALES field of view. These offsets were always applied while in the off-source nod position, and always along one axis at a time. The largest discontinuities in the x- and y-positions correspond to shifts of 0.1$\arcsec$ along the given axis. Neglecting the discontinuities, the drift of the stellar PSF in both the x- and y-directions follow arcs with turn-overs approximately 4.5~hours into the observing sequence. This slow drift is correlated with the pointing altitude of the telescope and arises from flexure of the ALES lenslet array as the telescope rotates. As the observations were obtained in pupil-stabilized mode such that the field of view was rotating over time, the change in position of the companion throughout the data has an additional rotational component compared to the star. The drift arising from the flexure of the lenslet array therefore instead produces an inflection point $\sim$4.5~hours in the case of the companion.

\begin{figure*}
	\includegraphics[scale=0.77]{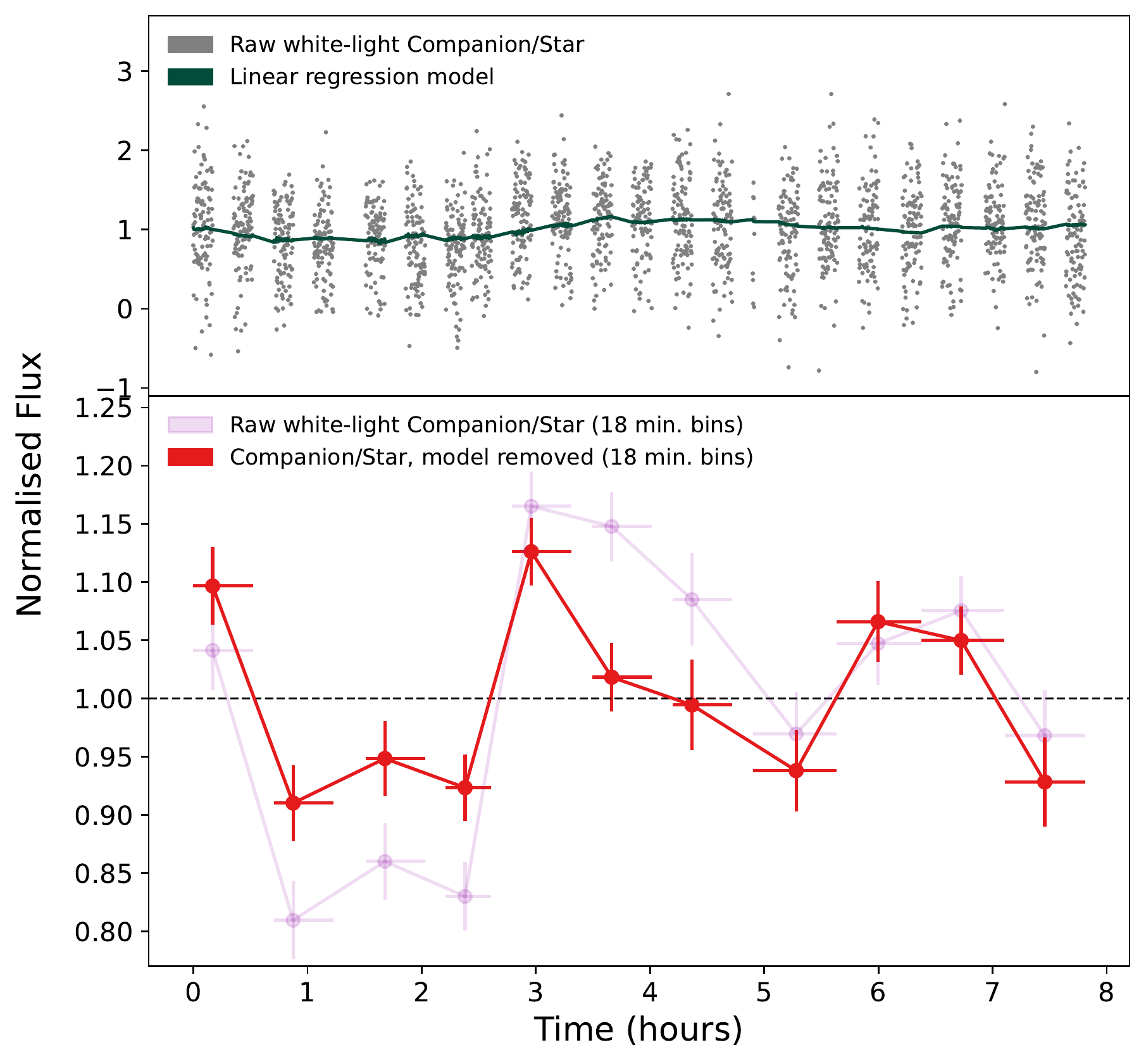}
    \caption{The model produced by the linear regression using the decorrelation parameter coefficients (in Table~\ref{table:lin_reg_coeffs}) is shown in the top panel in green alongside the raw differential light curve in grey. The bottom panel then shows in red the light curve produced when the raw differential light curve is divided by the linear regression model to remove the modelled trends that are not shared by the stellar and companion fluxes, binned to 18 minutes of integration time per bin. The raw differential light curve (i.e. prior to detrending) is also shown faintly in purple, reproduced for comparison from the bottom panel of Figure \ref{fig:raw_lcs}.}
    \label{fig:lin_reg_lc}
\end{figure*}

\begin{figure*}
    \centering
  \textcolor{white}{\frame{\includegraphics[width=0.33\textwidth]{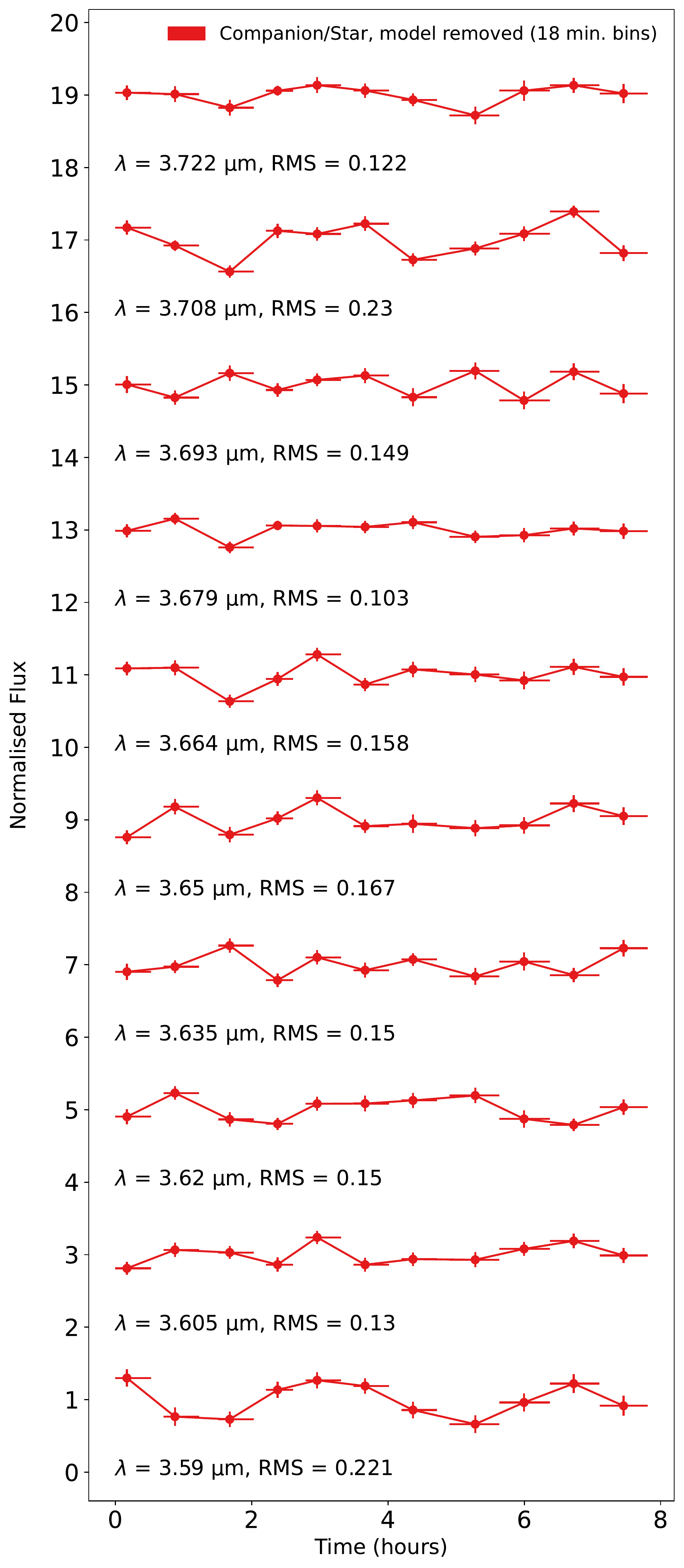}}}%
  \textcolor{white}{\frame{\includegraphics[width=0.33\textwidth]{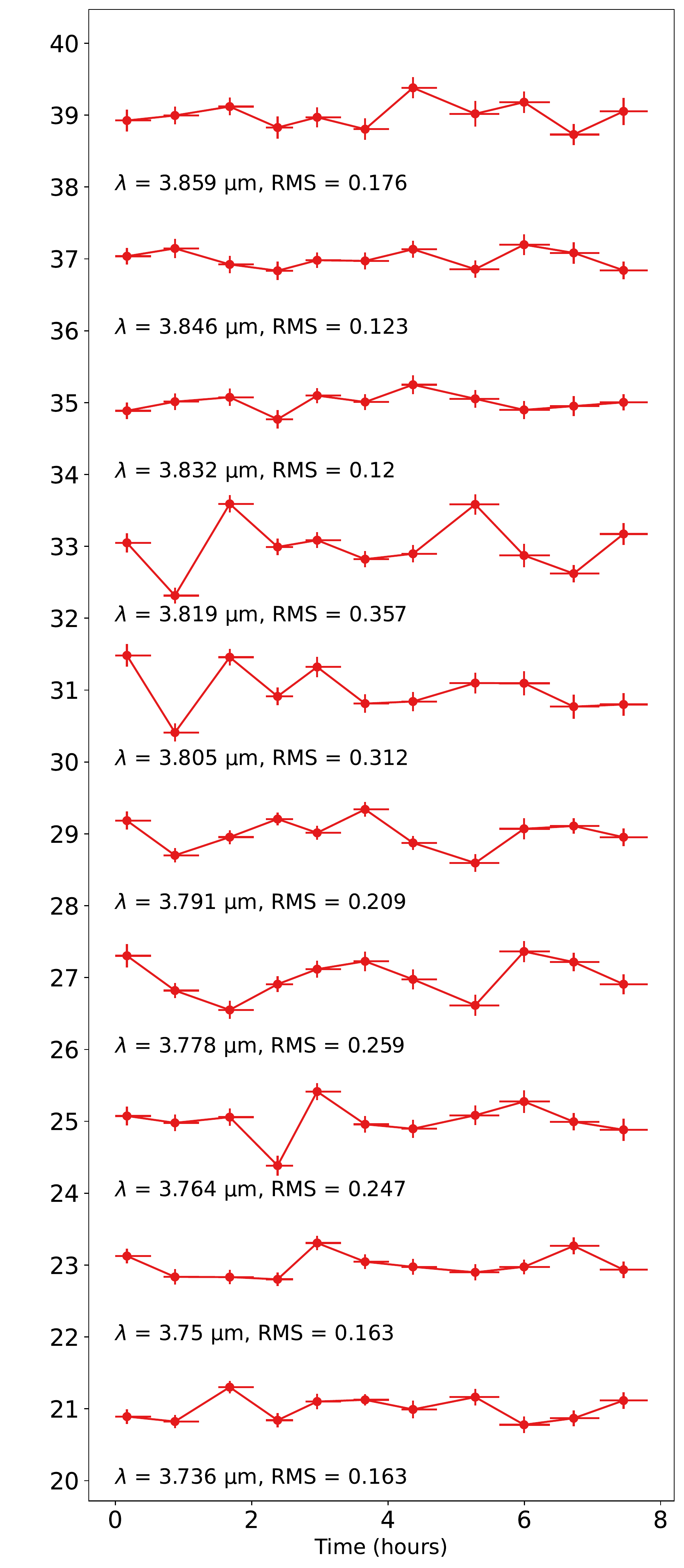}}}%
  \textcolor{white}{\frame{\includegraphics[width=0.33\textwidth]{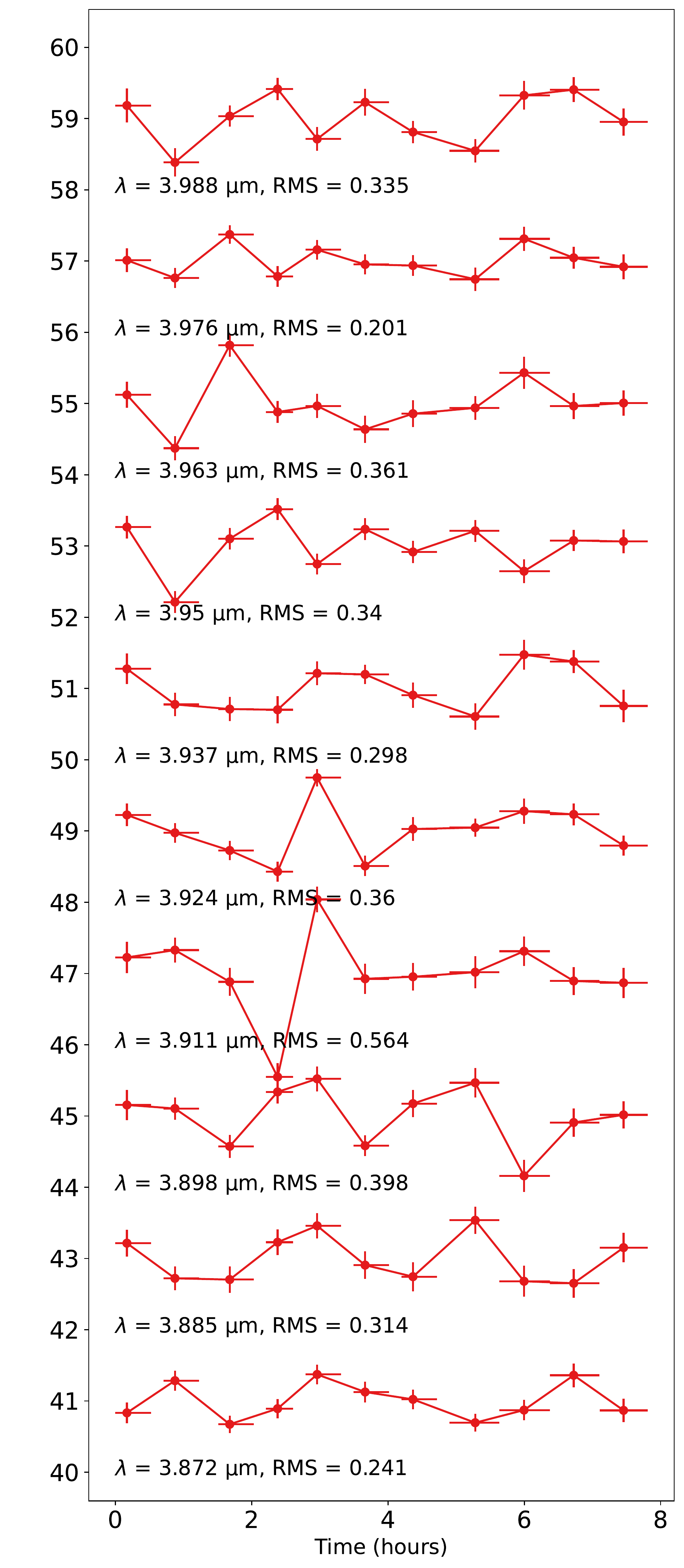}}}%
  \caption{In red, we show the detrended versions of the individual differential light curves in each of the 30 wavelength channels that were combined to produce the white-light curve. These wavelength channels cover $\lambda=$3.59-3.99~\textmu m, and are binned to 18 minutes of integration time per bin. An offset factor of 2 has been applied between each light curve to separate them from each other. Overall variability appears to increase with longer wavelength.}
  \label{fig:lc_30_plot}
\end{figure*}

We used the linear regression tools in version 1.0.2 of the scikit-learn Python package \citep{scikit-learn} to simultaneously fit these input parameters and produce a model fit to our differential light curve. The coefficients of the linear model produced by the linear regression are shown in Table~\ref{table:lin_reg_coeffs}, and the model itself is shown in green in the top panel of Figure \ref{fig:lin_reg_lc}, relative to the raw differential white-light curve in grey. We then divided the raw differential light curve by this model to produce a detrended version, shown in red in the bottom panel of Figure \ref{fig:lin_reg_lc}. We also overplot the raw differential light curve from the bottom panel of Figure \ref{fig:raw_lcs}, prior to detrending, in purple to allow the two to be compared. We also repeated this detrending process to produce detrended differential light curves for each of the 30 individual wavelength channels over the 3.59-3.99~\textmu m wavelength range that comprise the white-light curve. These are shown in Figure \ref{fig:lc_30_plot}, again binned to 18 minutes. We note that while ALES has a resolution of R$\sim$40, the raw data were spectrally extracted into 100 wavelength channels and so there is some correlation between wavelength channels.

\begin{flushleft}
\begin{table}
\caption{The decorrelation parameters $x_{i}$ used for the linear regression and the corresponding coefficients $c_{i}$ and intercept $c_{0}$ of the resulting linear model fit to the raw differential light curve. The linear model fit is then given by $y = (\sum_{i=1}^{n}c_{i}x_{i}) + c_{0}$.}
\begin{tabular}{p{0.41\columnwidth}p{0.2\columnwidth}}
\hline
Parameter ($x_{i}$)&Value ($c_{i}$)\\
\hline
Airmass&0.28782524\\
Air temperature&0.10690617\\
Star x-position&0.04819219\\
Star y-position&-0.04368692\\
Companion x-position&-0.02608288\\
Companion y-position&0.02129148\\
Wind speed&0.00059976\\
Wind direction&0.00050927\\
\hline
Intercept ($c_{0}$)&-1.318553799\\
\hline
\end{tabular}
\label{table:lin_reg_coeffs}
\end{table}
\end{flushleft}

\section{Results}\label{results}
The detrended differential white-light curve (Figure \ref{fig:lin_reg_lc}) shows sinusoidal-like variability over a timescale of a few hours. In the binned light curve, the normalised flux ranges from a minimum of 0.91 at 0.874 hours to a maximum of 1.13 at 2.961 hours. To better allow us to estimate the differential precision that we achieve in our light curve, we fitted the variability and removed it from our light curve. As the variability signal appears periodic, we used the NASA Exoplanet Archive periodogram service\footnote{\url{https://exoplanetarchive.ipac.caltech.edu/cgi-bin/Pgram/nph-pgram}} to apply the Lomb-Scargle algorithm to the unbinned detrended differential white-light curve and thereby search for sinusoidal periodic signals \citep{1976Ap&SS..39..447L, 1982ApJ...263..835S}.

\begin{figure}
	\includegraphics[width=\columnwidth]{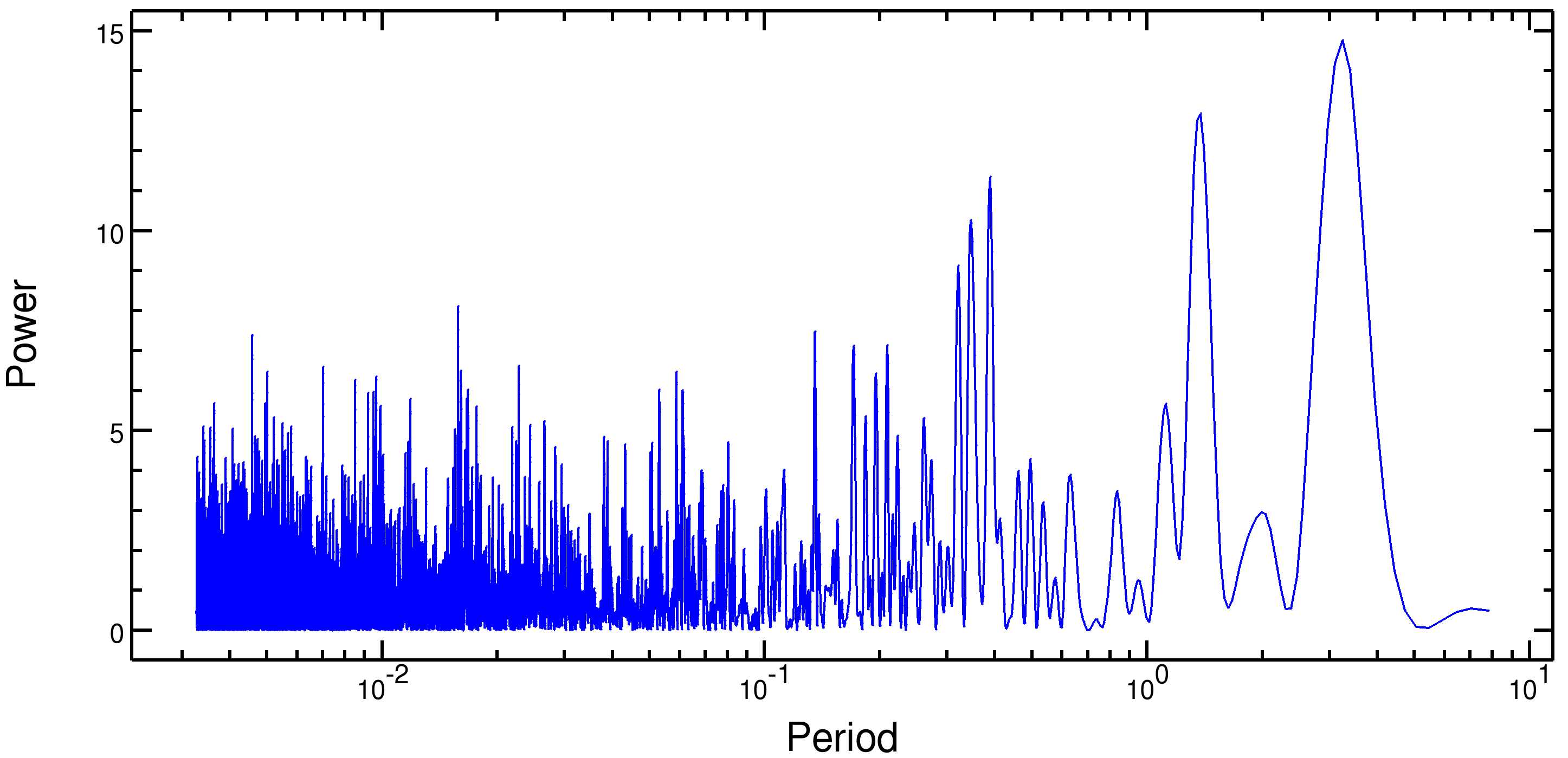}
	\includegraphics[width=\columnwidth]{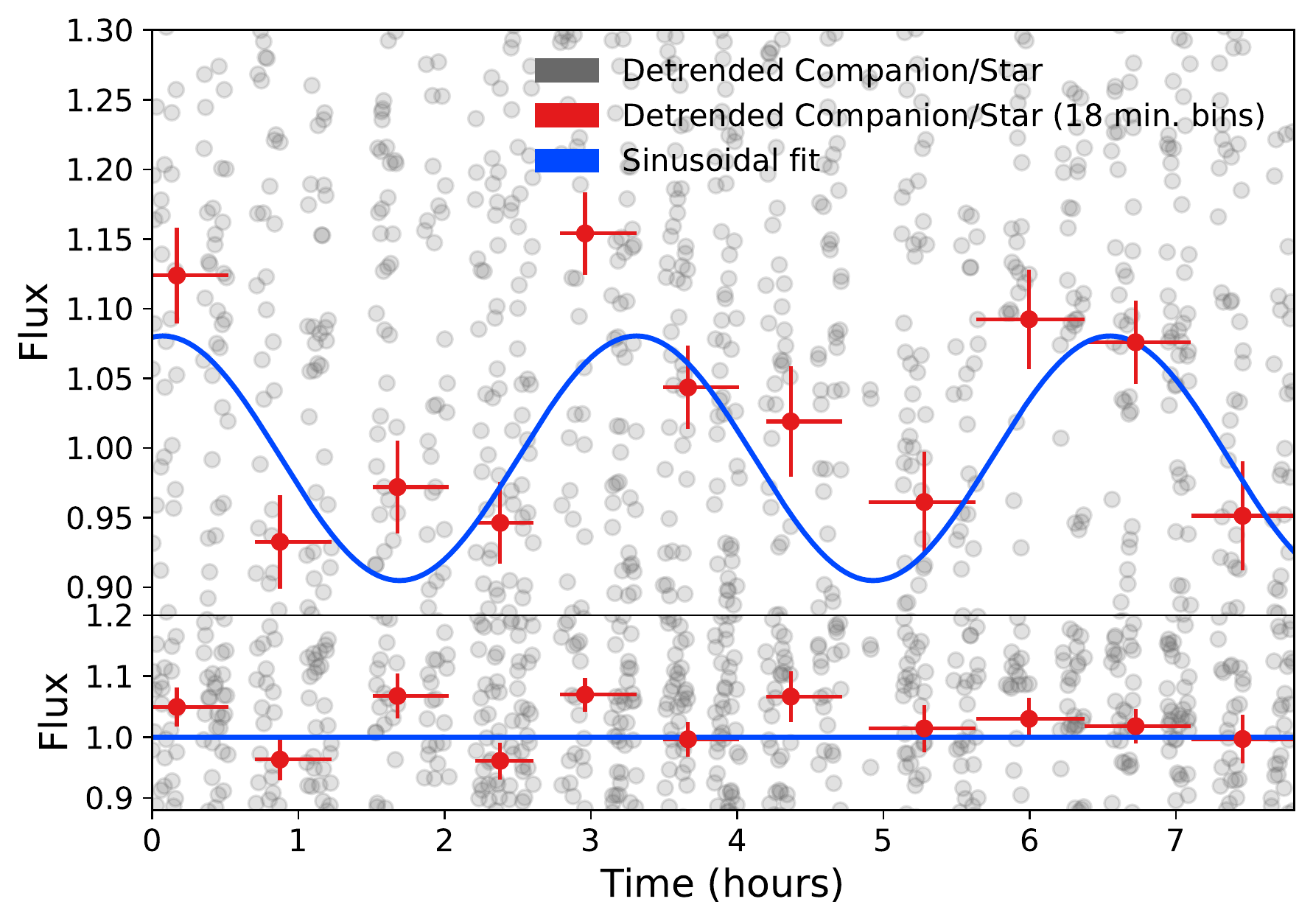}
	\includegraphics[width=\columnwidth]{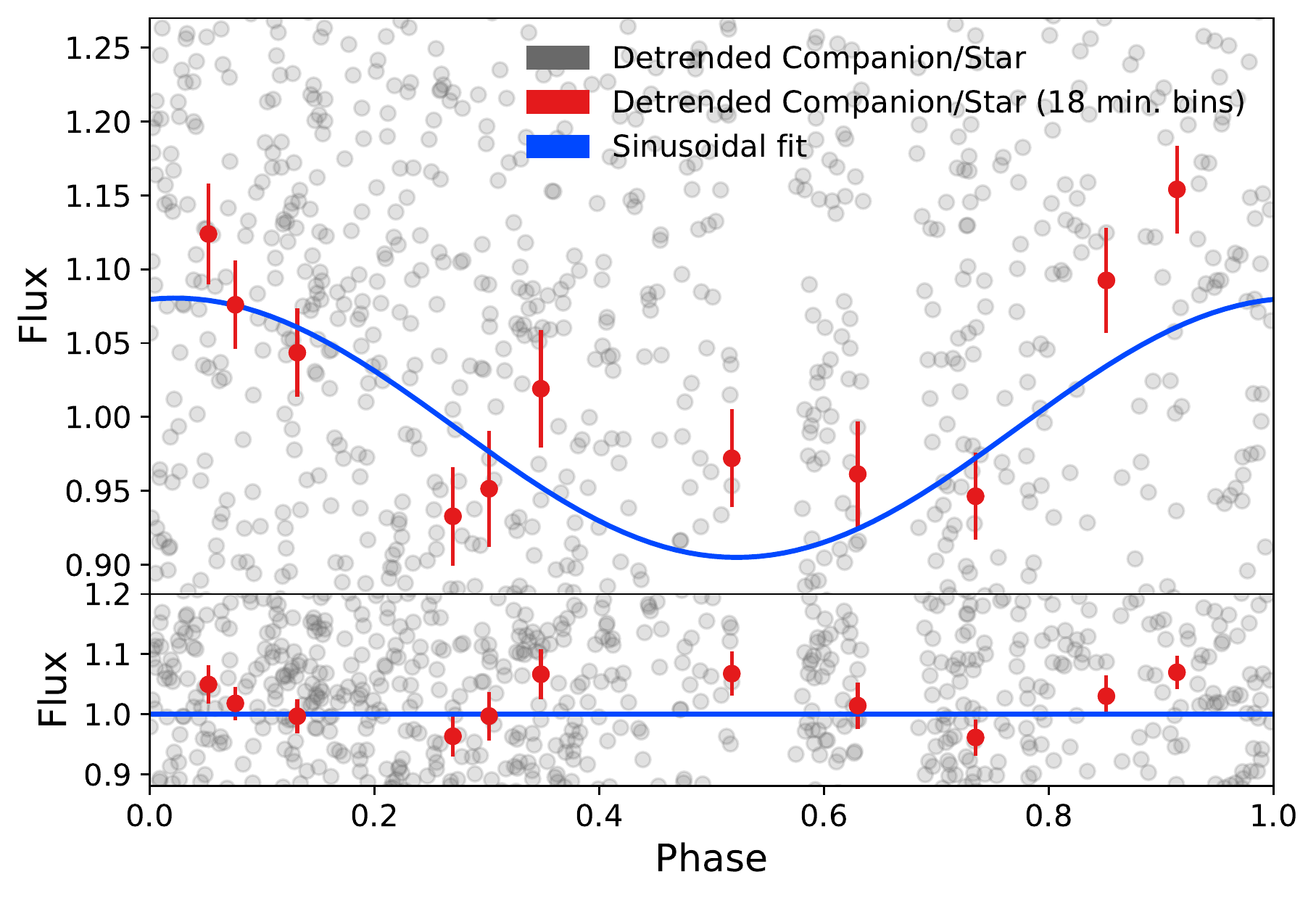}
    \caption{Top panel: the Lomb-Scargle periodogram for the unbinned detrended differential white-light curve. The strongest peak is at a period of 3.242~hours. Centre panel: the same detrended differential white-light curve from the bottom panel of Figure~\ref{fig:lin_reg_lc}, unbinned in grey and binned to 18~minutes of integration time per bin in red. The blue line is the fitted sinusoid with a semi-amplitude of 0.088 and a phase shift of 0.228. The residuals when the fitted sinusoid is divided out from the light curves are shown underneath. Bottom panel: the same as the panel above, but phase-folded to a period of 3.24~hours.}
    \label{fig:sin_plot}
\end{figure}

\begin{figure}
	\includegraphics[scale=0.46]{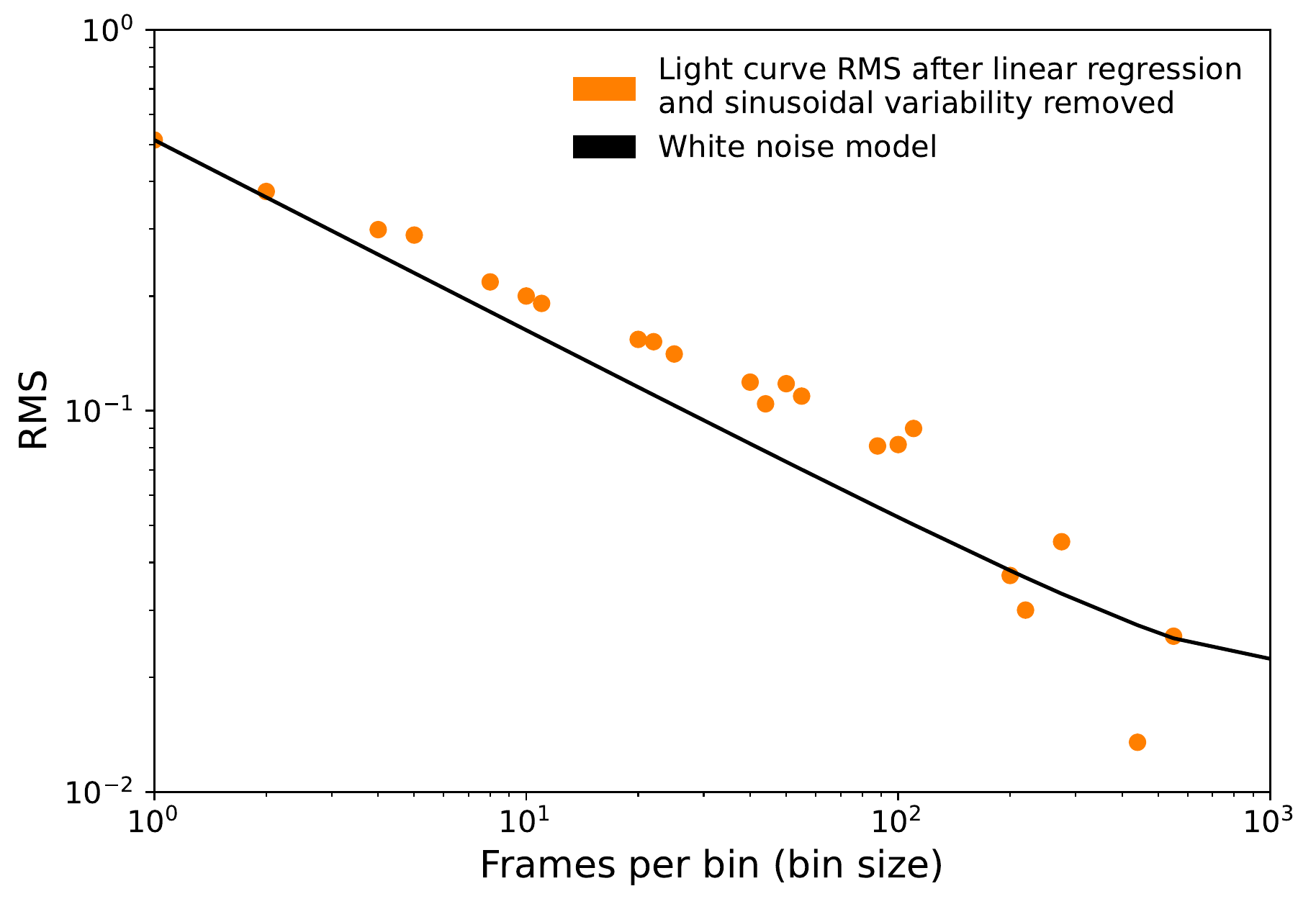}
    \caption{The RMS of the binned detrended differential white-light curve, after removing sinusoidal variability, is shown in orange as a function of bin size. The black line shows the theoretical white noise model, or the expectation of independent random numbers for a given bin size.}
    \label{fig:rms_plot}
\end{figure}

The strongest peak in the resulting periodogram (top panel, Figure~\ref{fig:sin_plot}) is at a period of 3.242~hours. We then fit a sinusoid to the light curve using this period as an initial guess, which returned a function with the same period (3.239), a semi-amplitude of 0.088, a phase shift of 0.228, and a $y$-offset of 0.993. This sinusoid is shown overplotted on the detrended light curve in the centre panel of Figure~\ref{fig:sin_plot}. The differential light curve was then divided by the fitted sinusoid to remove the variability signal to the first order (centre panel residuals, Figure~\ref{fig:sin_plot}). The bottom panel of this figure is the same as the panel above but with the data phase-folded to the period of the fitted sinusoid. Next, we followed the method of \citet{2011ApJ...733...36K} to assess the degree of ``red'' (correlated) noise in our light curve. We binned our detrended differential white-light curve, with the sinusoid removed, to a range of bin sizes, before normalising and subtracting one to centre around zero. We then measured the RMS of each resulting binned light curve. These RMS values are plotted against bin size in Figure \ref{fig:rms_plot}, alongside the expectation of independent random numbers as a function of bin size i.e. the white noise. For our chosen binning of 18~minutes of integration time per bin, which has a bin size of 200 frames per bin, we find an RMS of 0.037. For comparison, the RMS of the detrended differential white-light curve prior to the removal of the sinusoid, but with the same binning, is 0.073. We therefore conclude that the light curve of HD~1160~B shows variations with a semi-amplitude of $\sim$8.8\% or peak-to-peak amplitude of $\sim$17.6\%, and that the differential precision achieved in the binned light curve is at the 3.7\% level. The amplitude of the variations is therefore above the measured precision. Furthermore, this estimate of the precision is likely a conservative one; the variability signal is unlikely to have been perfectly removed by the sine fit and so the measured RMS values may be higher than the true limiting precision. A caveat of this result is that the baseline of our observations is only $\sim$7.81~hours, so we only cover $\sim$2.4 periods. Additional data is therefore needed to confirm the periodicity and amplitude of the variability of HD~1160~B. We discuss these results further and compare the precision achieved to similar studies in the literature in Section \ref{lc_precision}.

\section{Discussion}\label{discussion}
In this section we discuss the relative impact of the decorrelation parameters on our results, and the physical explanations of the systematics they introduce. We also compare the precision that we achieve to other variability studies in the literature that use different techniques, and discuss the potential application of differential spectrophotometry in future work.
\subsection{Impact of decorrelation parameters}\label{impact_params}
Using the linear regression coefficients of each parameter from Table~\ref{table:lin_reg_coeffs}, we can assess which parameters have the greatest impact on the light curve of HD~1160~B. The small angular separation of the companion and star might suggest that the effect of airmass would be small. Airmass can be an important systematic for differential spectrophotometric observations where other stars are used as the simultaneous photometric reference as the angular separation between the target and the reference can be large, causing the light from the two objects to pass through different atmospheric columns \citep[e.g.][]{2005AN....326..134B, 2022MNRAS.510.3236P}. However, in our case we use the companion's host star as the photometric reference, so the angular separation between the two is much smaller than is generally the case for observations using reference stars in the field. However, there is a significant colour difference between the star and the companion, which leads to different degrees of extinction at a given airmass. Therefore, we expect an airmass dependence and indeed it has the largest coefficient in the detrending model. Similar extinction effects are often seen in studies of transiting exoplanets, and can also exhibit a non-linear wavelength dependence \citep[e.g.][]{2022MNRAS.510.3236P}.

We also find the air temperature at LBT to be one of the parameters that is most correlated with our differential light curve. As the air temperature changes, this can potentially cause slight changes in the optical path of the telescope or instrument that lead to this correlation.

We further predicted that the positions of the companion and the star could introduce significant non-shared systematics to our measurements; as HD~1160~A was not pixel-stabilised during our observations, we are sensitive to both intra- and interpixel variations as the star and the companion move across the detector. Both the star and the companion changed position on the detector due to flexure of the ALES lenslet array as the instrument moved and positional offsets applied intentionally to keep the target close to the centre of the field of view. We also nodded on and off of the target to enable background subtraction. While the process of nodding is itself relatively accurate, it is not repeatable at a sub-pixel pointing precision, introducing a slight offset error between nods.  Furthermore, LBTI data is always pupil-stabilized, so the field of view was rotating throughout the night. Although it may have been optimal to fix the star and companion positions to the same detector pixels for the duration of the observations, this was not possible due to the effect of the lenslet array flexure and the lack of instrument derotator in the LBTI architecture \citep{2022AJ....163..217D}. However, we find that these positional changes are not the most correlated with our differential light curve. It is possible that this is because the light from the star and companion is spread out across multiple detector pixels when spectrally dispersed, reducing the impact of any systematic issues arising from any single pixel. The light of our target is dispersed across wavelength, similar to a technique often used for high-precision differential photometry whereby a telescope is intentionally defocused to disperse starlight over the detector \citep[e.g.][]{2011A&A...528A..49D, 2013A&A...550A..54D, 2012ApJ...760..140C, 2015ApJ...802...28C}. This has the effect of reducing systematics due to intrapixel variations and minimises the impact of any residual flat-field errors. The step of recombining our data into a white-light curve may therefore have helped to reduce systematic trends that would otherwise have been introduced by the positional movement of the targets throughout the night. The use of a spectrograph also has the additional benefit of allowing us to remove individual wavelength channels that are found to contain defects. For this dataset, this allowed us to leave out channels affected by overlapping spectral traces, significant telluric absorption, and absorption by the glue layer of the dgvAPP360, as described in Section \ref{channel_selection}.

Lastly, we find that the wind speed and direction are the least correlated with our differential light curve, but note that the conditions on the night were exceptionally stable and so cannot rule out that these factors could have an impact in less optimal conditions.

For future observations, residual systematics in differential light curves of directly imaged companions could be removed using more advanced methods from the exoplanet transmission spectroscopy and high-precision secondary eclipse literature such as fitting the data using a Gaussian process regression \citep[e.g.][]{2012MNRAS.419.2683G, 2017Natur.548...58E, 2018MNRAS.474.1705N, 2018Natur.557..526N, 2020AJ....160...27D, 2020AJ....160..188D, 2021MNRAS.503.4787W}. In general, methods used to identify trends in transmission spectroscopy data also include a model of the exoplanet transit itself, allowing the transit to be detected even when the strength of the signal is very low. However, this is possible because the expected shape of the signal is well understood. This is more difficult for searches for variability in directly imaged companions, where the expected shape of the variability signal is not necessarily well-known in advance. Furthermore, many of these methods assume a linear relationship between systematics and trends in the light curve, while the telluric and instrumental systematics present in time-series data can be complex and non-linear. In the future, an optimal approach should account for the functional form of the correlated parameters \citep{2022MNRAS.510.3236P, 2022MNRAS.515.5018P}.

\subsection{Differential light curves}\label{lc_precision}
\subsubsection{Variability interpretation}\label{var_causes}
In Section \ref{results} we found that HD~1160~B shows variations with a semi-amplitude of $\sim$8.8\% (or peak-to-peak amplitude of $\sim$17.6\%). To compare this result to literature observations of similar objects, we must consider both spectral type and the wavelengths at which variability has been observed. \citet{2022ApJ...924...68V} found that virtually all L dwarfs are likely to be variable at the 0.05-3\% range, and several studies have measured higher variability, up to 26\% \citep[e.g.][]{2012ApJ...750..105R, 2014ApJ...797..120R, 2016ApJ...829L..32L, 2018AJ....155...95B, 2020ApJ...893L..30B}. However, there is evidence that brown dwarf variability amplitude may have a strong wavelength dependence. For example, HST observations of highly variable L dwarf companion VHS~1256-1257~b identified a large variability amplitude of 24.7\% at 1.27~\textmu m, while Spitzer observations at 4.5~\textmu m found a far lower amplitude of 5.76$\pm$0.04\% \citep{2020ApJ...893L..30B, 2020AJ....160...77Z, 2022arXiv220900620M}. \citet{2020AJ....160...77Z} do note, however, that the HST and Spitzer observations were not obtained contemporaneously and so the atmosphere of VHS~1256-1257~b and hence its variability properties are likely to have changed substantially over the intervening timescale. Comparisons of large surveys also suggest that variability amplitudes are lower in the mid-infrared than in the near-infrared, although there is evidence for a weaker wavelength dependence and enhanced mid-infrared variability amplitudes for the young isolated brown dwarfs most similar to substellar companions \citep[e.g.][]{2014ApJ...793...75R, 2015ApJ...799..154M, 2018AJ....155...95B, 2022ApJ...924...68V}.

The amplitude of the L-band variability we measure for HD~1160~B is quite extreme compared to literature results, but it is important to note that there have been very few variability studies for substellar companions in the mid-infrared, making direct comparison difficult. High-amplitude variability in brown dwarfs is generally attributed to heterogeneous surface features, such as spots or clouds of varying thickness, rotating in and out of view as the object rotates \citep[e.g.][]{2013ApJ...768..121A, 2017AstRv..13....1B, 2018haex.bookE..94A}. Some light curves show more complex features that cannot be modelled with a single atmospheric feature, or features that evolve over short or long timescales \citep[e.g.][]{2009ApJ...701.1534A, 2015ApJ...799..154M}. These features and time evolution may arise from changing weather systems, or bands of clouds which rotate within the target's atmosphere and generate waves on a global scale \citep[e.g.][]{2017Sci...357..683A, 2021MNRAS.502..678T}. For HD~1160~B, observations over a longer baseline are required to be able to characterise any time evolution in the variability signal.

However, the spectral type of HD~1160~B is unclear as it has a highly peculiar spectrum that cannot be satisfactorily fit with spectral models or templates in current libraries, and some studies suggest that it could instead be a late-M dwarf \citep{2016A&A...587A..56M, 2017ApJ...834..162G, 2020MNRAS.495.4279M}. If HD~1160~B is an M dwarf, its variability would most likely arise from cool star spots caused by magnetic activity in its photosphere, which are common in M-dwarfs and would rotate in and out of view in much the same way as the cloud features of lower mass objects \citep[e.g.][]{2002AN....323..333B, 2009A&A...508.1313F, 2009MNRAS.400.1548S, 2012MNRAS.427.3358G, 2021MNRAS.500.5106G, 2021MNRAS.504.4751J}. The properties of such variability can be highly dependent on the spot distribution and fractional spot coverage of a given object; some M-dwarfs have a very high coverage with multiple starspots covering as much as 20-50\% of their fractional surface area inhomogeneously \citep[e.g.][]{2004AJ....128.1802O, 2010ApJ...718..502M, 2011ApJ...742..123I, 2012MNRAS.427.3358G, 2013MNRAS.431.1883J}. Spot-induced variability amplitudes for M-dwarfs generally range from the subpercent level up to around $\sim$5\% \citep[e.g.][]{2006A&A...448.1111R, 2009Natur.462..891C, 2012MNRAS.426.1507B, 2012A&A...538A..46D, 2013MNRAS.431.3240N}. Although observed far less often in the infrared compared to the optical, flaring events can induce far stronger variability in M-dwarfs at amplitudes ranging from the subpercent level up to tens of percent \citep[e.g.][]{2012MNRAS.427.3358G, 2012AJ....143...12T}. Our measured variability amplitude for HD~1160~B is therefore also on the higher end of what has been observed for earlier spectral types such as late M- and early L-dwarfs, barring flares, although we again note the lack of literature studies of similar objects in the L-band.

While the variability observed for HD~1160~B appears high, another point to consider is its orbital inclination, which the latest orbital fits suggest is close to edge-on as viewed from Earth \citep[92.0{\raisebox{0.5ex}{\tiny$\substack{+8.7 \\ -9.3}$}}\textdegree;][]{2020AJ....159...63B}. If the obliquity of HD~1160~B is aligned with its orbit such that we are viewing its rotation close to edge-on, the observed variability amplitude is likely to compose a much larger fraction of its true variability compared to if it were viewed face-on. Indeed, \citet{2017ApJ...842...78V} demonstrated that the highest variability amplitudes are seen for targets with close to edge-on viewing angles.

If we interpret the $3.24$ h sinusoidal variation we observed as the true rotation period of HD~1160~B, we can further consider this within physical limitations. The breakup period of a rotating object is dependent on its radius, which is itself age-dependent, and mass. Both age and mass are poorly constrained for HD~1160~B: literature results place the system's age in the 10-300~Myr range \citep{2012ApJ...750...53N, 2016A&A...587A..56M, 2019AJ....158...77C}, and mass estimates range from $\sim$20~M\textsubscript{Jup} to 123~M\textsubscript{Jup} \citep{2019AJ....158...77C, 2020MNRAS.495.4279M}. \citet{2020AJ....160...38V} calculated the breakup periods of brown dwarfs as a function of age by equating equatorial velocity with the escape velocity, accounting for radial contraction over time. When we compare our measured HD~1160~B variability period of 3.24~hours to their results (their Figure 13), we find that this is a physically feasible rotation period for most possible combinations of mass and age from the literature, albeit very close to the breakup period in many cases. An alternative explanation is that the 3.24~hour variability signal that we see is produced by multiple features in the atmosphere of HD~1160~B, and that its rotation period is actually longer \citep{2016ApJ...830..141L}. Additional observations of this variability over a longer baseline will help to further characterise its origin and confirm whether its periodicity reflects that of the companion's rotation.

The detrended differential light curves for each of the 30 individual wavelength channels over the 3.59-3.99 µm wavelength range that comprise the white-light curve (Figure \ref{fig:lc_30_plot}) show increasing statistical errors at longer wavelengths, as expected as our S/N is lower here. Using the RMS of each light curve as a metric to compare the scatter in each channel, we do see tentative evidence of increasing variability towards longer wavelengths beyond the increase of RMS expected from the S/N. However, as the total baseline of our observations is only a single night, we see too few repetitions to be confident of variability patterns in individual wavelength channels. Additional spectrophotometric data will therefore be required to confirm this. Although we modelled the overall variability in our white-light curve with a single sinusoid, it is also possible that the phase and amplitude is different per wavelength channel as distinct atmospheric features at separate locations in the atmosphere of the companion may produce variability with a different wavelength dependence. Furthermore, although the overall wavelength range of these 30 channels is relatively small, different wavelengths probe different pressure levels in the companion's atmosphere, and hence different layers of the atmosphere \citep[e.g.][]{2012ApJ...760L..31B, 2013ApJ...778L..10B, 2017Sci...357..683A, 2019AJ....157...89G}.

\subsubsection{Light curve precision}\label{precision}
In Section \ref{results}, we found that after removing a single sinusoid from the data, we achieved a precision of 3.7\% in the detrended differential light curve when it is binned to a bin size of 200 data points, corresponding to 11 bins of 18 minutes of integration time. There have been three previous studies searching for variability in substellar companion from the ground; \citet{2016ApJ...820...40A}, \citet{2021MNRAS.503..743B}, and \citet{2022AJ....164..143W} each conducted variability searches on the HR~8799 planets using satellite spots as photometric references. In a pilot variability study, \citet{2016ApJ...820...40A} reach a $\sim$10\% planet-to-planet photometric accuracy for SPHERE observations of 25 minute cadence when data from different nights are combined for a total telescope time of 3.5~hours. \citet{2021MNRAS.503..743B} goes further with SPHERE to conduct a longer (>4~hours) search, successfully constraining the sensitivity to variability to amplitudes $>$5\% for HR~8799b and $>$25\% for HR~8799c. More recently, \citet{2022AJ....164..143W} used SCExAO/CHARIS to improve the variability constraints of HR~8799c to the 10\% level, and HR~8799d to the 30\% level. They did this by combining the use of satellite spots with a spectrophotometric approach similar to the one we present in this paper, using the CHARIS IFS to disperse the light into individual spectra before recombining the channels into wider bands. At first glance, the sensitivity that we achieve for HD~1160~B here may appear to compare favourably with these results. However, it is important to consider a number of caveats that make direct comparison unjustified. All three of the HR~8799 studies were conducted at near-infrared wavelengths with 8.2-m telescopes, while ours was in the mid-infrared with an 8.4-m telescope. More significantly, the HR~8799 planets are fainter than HD~1160~B, with contrasts of $\Delta H=8-10$ mag compared to their host star, which has a H-band magnitude of 5.28 mag \citep{2008Sci...322.1348M, 2010Natur.468.1080M}. HD~1160~B is brighter, with a contrast of $\Delta L^{\prime}=6.35$ mag compared to a host star with an $L^{\prime}$-band magnitude of 7.06 mag \citep{2012ApJ...750...53N}. The lower sensitivity to variability achieved by these studies is therefore partially a reflection of the intrinsically lower fluxes of their targets, which leads to higher errors on their photometry.

However, each of the HR~8799 variability studies also found that the satellite spots can demonstrate individual variations of their own and are often anti-correlated with each other. This means that they may not always serve as appropriate photometric references with which to detrend the light curve of a companion. \citet{2022AJ....164..143W} found that the flux ratio of the SCExAO satellite spots shows time variation with a scatter of $\sim$3\% across a night, and can show even larger variations on a shorter timescale, up to 10\%. This potentially sets a limit to the precision that can be achieved using satellite spots, particularly on nights where observing conditions are less stable.

A key advantage of the dgvAPP360 compared to satellite spots is its simplicity; the photometric reference it provides is simply an image of the host star, and so it does not suffer from the same correlated systematics as the satellite spots. Differential photometry between the companion and the star can be carried out directly. It may be possible to reach an even deeper precision through differential spectrophotometry with a dgvAPP360 than the 3.7\% level that we achieve here. Indeed, if we compare the detrended light curve RMS as a function of bin size to the white noise expectation (Figure \ref{fig:rms_plot}), it continues to follow the trend of the white noise and does not plateau implying that we have not yet reached any noise floor. This means that in principle the precision of the differential light curve would improve further if more data from additional epochs was added. This also indicates that this technique should remain usable for companions with less favourable contrasts than HD~1160~B, such as those in the planetary-mass regime. However, more data per bin will be required to achieve the same precision for a fainter companion, so the time-sampling in the binned light curves may be less fine in these cases.

Many transiting exoplanet studies make use of a region of the target light curve that is expected to be flat (i.e. an out-of-transit baseline) to test the degree to which systematics have been corrected. While we have shown here that key systematic trends are successfully removed in the detrended differential white-light curve of HD~1160~B, we do not have such a baseline to verify the level of impact of any remaining systematics. The possibility therefore remains that an unknown systematic could be present that has not been accounted for by any of the processes that we have applied here, and could be responsible for the variability that we see in the light curve of HD~1160~B. However, this is also inherently the case for any study that explores the variability of isolated brown dwarfs and planetary-mass objects, stellar variability due to star spots or other sources, or transiting exoplanet studies where exoplanets transit variable stars \citep[e.g.][]{2002ApJ...577..433G, 2006A&A...448.1111R, 2013ApJ...778L..10B, 2015ApJ...813L..23B, 2021MNRAS.503..743B, 2013ApJ...767...61G, 2014ApJ...793...75R, 2014A&A...566A.111W, 2017AJ....154..138N, 2019A&A...629A.145E, 2019MNRAS.483..480V, 2021AJ....162..179M, 2022AJ....164...65M}. A subsequent study is forthcoming in which we further investigate the precision that can be reached with this technique, and use injection-recovery tests to assess the extent to which known, simulated variability signals can be recovered (Sutlieff et al., in preparation).

Nonetheless, ground-based differential spectrophotometry with the vAPP is highly complementary and advantageous to space-based approaches for measuring the variability of high-contrast companions. There have been many successful space-based measurements of companion variability using HST, detecting variability with amplitudes down to the 1-2\% level in some cases \citep[e.g.][]{2018AJ....155...11M, 2019ApJ...875L..15M, 2020ApJ...893L..30B, 2020AJ....159..140Z, 2020AJ....160...77Z}. \citet{2016ApJ...818..176Z} was further able to detect sub-percent variability using HST observations of planetary-mass companion 2M1207b, which lies at roughly the same angular separation as HD~1160~B, albeit with a more favourable contrast. Furthermore, the first variability monitoring with JWST, which should reach an even greater precision, is currently underway as part of the Early Release Science Program \citep{2022PASP..134i5003H}. However, while JWST has the sensitivity to image fainter, lower mass companions and measure their variability with great precision, its $\sim$6.5~m mirror is smaller than those of the largest ground-based telescopes, and thus is cannot outperform large ground-based telescopes with extreme AO at small separations $\lesssim0.5\arcsec$ at $\sim$3.5~\textmu m \citep{2022SPIE12180E..3QG}. This means that companions at the closest angular separations such as Jupiter analogues are for now likely only accessible with ground-based monitoring techniques, for all but the nearest stars \citep{2021MNRAS.501.1999C, 2022SPIE12180E..3NK}. Ground-based telescopes also uniquely provide access to higher resolution spectrographs, such that line profile variability could be used in Doppler imaging to create 2D global maps of features in exoplanet atmospheres such as storms similar to Jupiter's Great Red Spot \citep[e.g.][]{2014A&A...566A.130C}.

\subsection{Observing strategy}\label{obs_strat}
During the observing sequence, we obtained 6 wavelength calibrations at intervals throughout the night to allow us to test whether differences in the wavelength calibration used would lead to differences in the differential light curve. In Figure \ref{fig:lc_wcal_comparison} in Section \ref{results_wavecal}, we found that the differential light curve is robust to changes in the wavelength calibration. It is therefore preferable to acquire wavelength calibrations at the start or end of future observations, perhaps along with a single precautionary wavelength calibration at high elevation, and instead obtain additional data on target and minimise pixel offsets.

We also used an on/off nodding pattern to enable background subtraction. However, future studies may wish to consider alternative methods to remove the thermal background such that the amount of time spent on target can be doubled and the entire system stabilised. For example, \citet{2022AJ....163..217D} developed an approach whereby 93\% of the frames obtained were on-target and the thermal background was modelled and removed using the science frames themselves and a small number of background frames obtained before and after the observing sequence. As Figure \ref{fig:rms_plot} shows that we approach the photon noise limit, increased on-target time would therefore allow a greater differential light curve precision to be obtained in a single night of observations.

Lastly, the absence of an instrument derotator in the LBTI architecture meant that the field of view was rotating throughout the observing sequence. In addition to the drift due to lenslet flexure and the manual offsets applied to keep HD~1160~A centred in the field of view, this meant that HD~1160~A and B were not pixel-stabilised during our observations. However, the linear regression correlation coefficients of the positions of the star and the companion are small relative to those of airmass and air temperature. This suggests that that, when present, these factors dominate over any effect from HD~1160~A and B not being pixel-stabilised.

Understanding whether the host star of a given target is itself varying is important when interpreting the trends in a differential light curve. Most, if not all, potential targets for differential spectrophotometry will be present in at least the TESS full frame images available on the MAST archive. Even though this data will most likely not be contemporaneous with a particular set of observations, the total baseline of the coverage should be relatively long and therefore sufficient to check a host star for variability at the required precision, especially if the target appears in multiple TESS sectors. We therefore recommend this method as a good way to verify the level of variation shown by the host star of a target for differential spectrophotometry, and hence whether it is stable enough to act as a simultaneous photometric reference without requiring further analysis to account for stellar variability.

\subsection{Outlook}\label{outlook}
In principle, the technique presented in this paper can be applied to any vAPP coronagraph used in combination with an IFS. Although ALES is currently the only IFS operating over the L and M bands \citep{2021ApOpt..60D..52D}, a vAPP coronagraph is also available on SCExAO/CHARIS on the 8.2-m Subaru Telescope, offering R$\sim$19 spectrographic coverage over the J, H, and K bands (1.13-2.39~\textmu m) \citep{2016SPIE.9908E..0OG, 2017SPIE10400E..0UD, 2019A&A...632A..48B, 2020SPIE11448E..7CL, 2021A&A...646A.145M}. There will also be two different vAPPs on the Mid-infrared ELT Imager and Spectrograph (METIS) instrument on the upcoming 39-m Extremely Large Telescope (ELT), for which this work is a pathfinder. METIS will provide high spectral resolution spectroscopy (R$\sim$100,000) over the L and M bands \citep{2016SPIE.9909E..73C, 2018SPIE10702E..1UB, 2021Msngr.182...22B, 2018SPIE10702E..A3K}. Variability measurements using a vAPP may even be possible for broad-band imaging data where an IFS is unavailable, although sensitivity will be inherently more limited without the benefits of using differential spectrophotometry to reduce the effects of systematics. There are several vAPPs currently available on such coronagraphic imagers, such as MagAO \citep{2016SPIE.9909E..01M, 2017ApJ...834..175O, 2021MNRAS.506.3224S} and MagAO-X \citep{2019JATIS...5d9004M, 2020SPIE11448E..0UC} on the 6.5-m Magellan Clay Telescope, and the recently-commissioned Enhanced Resolution Imager and Spectrograph (ERIS) instrument on the VLT \citep{2018SPIE10702E..3YB, 2021JATIS...7d5001B, 2018SPIE10702E..46K, 2022MNRAS.515.5629D}, with more planned, including for GMagAO-X on the GMT \citep{2020arXiv200406808C} and MICADO on the ELT \citep{2018SPIE10703E..13C, 2018arXiv180401371P}. Using the vAPPs that will be available on larger telescopes, variability monitoring through differential spectrophotometry will be possible for fainter companions at closer angular separations, including those in the exoplanet mass regime. While this will be inherently more challenging for such companions at greater contrasts, these will remain accessible to this technique through the addition of data from multiple epochs as long as the systematic noise floor is not reached, albeit with a trade-off between light curve precision and time-sampling.

In the era of extremely large telescopes, high-contrast imaging combined with high resolution spectroscopy will provide access to fainter companions at lower masses and older ages and allow their orbital velocities and spin to be measured \citep[e.g.][]{2014Natur.509...63S, 2015A&A...576A..59S, 2016A&A...593A..74S, 2018arXiv180604617B, 2021AJ....162..148W, 2022ApJ...937...54X}. Measurements of how individual absorption lines change in depth and width as an exoplanet rotates will allow two-dimensional surface maps of exoplanet atmospheres and weather to be produced, through techniques such as Doppler imaging \citep[e.g.][]{2014A&A...566A.130C, 2014Natur.505..654C, 2021arXiv211006271L, 2022ApJ...933..163P}. Further in the future, multi-wavelength variability measurements obtained in reflected light may even enable exo-cartography of directly imaged Earth-like exoplanets \citep[e.g.][]{2019AJ....157...64L, 2022AJ....164....4L, 2020ApJ...894...58K, 2022ApJ...930..162K, 2022MNRAS.511..440T}.

A limitation of using the dgvAPP360 for variability measurements is that post-processing algorithms relying on angular diversity, such as ADI and PCA, cannot be used without also removing the central PSF of the star that we use as the simultaneous photometric reference. Furthermore, the stellar PSF must remain unsaturated throughout the observing sequence. This potentially limits the sample of targets with bright enough companions. Although HD~1160~B is bright enough that additional noise reduction techniques were not necessary to produce a detection of ample S/N, this may not be the case for fainter directly imaged companions in the exoplanet mass regime. However, it may be possible to use novel alternative algorithms to reach deeper contrasts.

For example, the Temporal Reference Analysis of Planets \citep[TRAP;][Liu et al., submitted]{2021A&A...646A..24S} algorithm instead relies on temporal diversity. TRAP reconstructs the systematics in a given region in the data using reference pixels that share the same underlying noise sources. By simultaneously fitting the model of a companion signal `transiting' over detector pixels and the light curves of the reference pixels, TRAP can then remove these systematics. It may be possible to leverage the information provided by TRAP to improve the companion S/N without removing the stellar PSF, or even to extract detrended light curves directly. Another option would be to use the gvAPP coronagraph, which is different from the dgvAPP360 in that it creates two images of the target star each with a 180\textdegree{} D-shaped dark hole on opposing sides, as well as an additional `leakage term' positioned between the two \citep{2012SPIE.8450E..0MS, 10.1117/12.2056096, 2021MNRAS.506.3224S}. The leakage term is an entirely separate PSF of the star that appears at a fraction of its full brightness, making it ideal as a simultaneous photometric reference and enabling observations of systems with host stars that would otherwise be too bright. Post-processing algorithms can be applied to the main PSFs to reach deeper contrasts without impacting the leakage term, enabling differential variability measurements provided that the impact of the algorithms on the photometry of the companion can be characterised precisely. However, the gvAPP coronagraph can suffer from wavelength-dependent smearing, which would make such measurements more complex than those obtained with a dgvAPP360 \citep{2017ApJ...834..175O}. In addition to the leakage term, some vAPPs (such as the VLT/ERIS gvAPP) produce other faint reference spots specifically for use as photometric references in situations where the core of the target star PSF is saturated \citep{2021ApOpt..60D..52D, 2022SPIE12184E..5MK}.

In addition to probing the intrinsic variability of high-contrast companions, differential spectrophotometry could also be used to observe the transits of satellites such as exomoons or binary planets passing in front of these companions \citep[e.g.][]{2016A&A...588A..34H, 2022MNRAS.516..391L}. Candidate satellites have been identified around transiting exoplanets, directly imaged companions, and isolated planetary-mass objects using a range of techniques, but none have yet been definitively detected \citep[e.g.][]{2018SciA....4.1784T, 2018A&A...617A..49R, 2019A&A...624A..95H, 2019ApJ...877L..15K, 2020A&A...641A.131L, 2020AJ....159..142T, 2021ApJ...918L..25L, 2021ApJ...922L...2V, 2022NatAs...6..367K}. For directly imaged companions, variability arising from transit events could be distinguished from that caused by inhomogeneous atmospheric features in similar ways to transiting exoplanets and star spots, by considering the companion light curves across the different wavelength channels. Transit signals are expected to be almost achromatic, while intrinsic variability is generally wavelength-dependent \citep{2019AJ....157..101M, 2021ApJ...918L..25L, 2023PASP..135a4401L, 2022MNRAS.516..391L}. \citet{2022MNRAS.516..391L} found using simulations that although the probability of successfully detecting smaller exomoons around a directly imaged companion is very low with current instrumentation and techniques, detections of larger binary planets are already within reach. New techniques to improve differential light curve precision for directly imaged companions, including differential spectrophotometry with the dgvAPP, will help to increase these probabilities further and potentially enable the first definitive detections of satellites around substellar companions.

\section{Conclusions}\label{conclusions}
We present a novel, ground-based approach for constructing differential light curves of high-contrast companions through direct differential spectrophotometric monitoring, using the double-grating 360\textdegree{} vector Apodizing Phase Plate coronagraph and the ALES integral field spectrograph. The dgvAPP360 allows high-contrast companions to be detected while also providing an image of the host star, which crucially can be used as a simultaneous photometric reference. We combine the dgvAPP360 with ALES to follow the highly successful technique of differential spectrophotometry used in exoplanet transmission spectroscopy, where light is spectrally-dispersed to reduce systematic effects that otherwise dominate the variability signal we aim to measure, and then recombined into white-light flux measurements.

We demonstrated this approach using a full night of observations of substellar companion HD~1160~B. The time-series fluxes of the companion and the star in each wavelength channel were extracted simultaneously using aperture photometry. We then produced white-light measurements for both the companion and the star at each time frame by taking the median combination of the photometry in the wavelength dimension. The companion flux was then divided by that of the star to eliminate trends common to both, arising from Earth’s atmosphere and other systematics, producing a differential white-light curve that only contains non-shared variations and covers a wavelength range of 3.59-3.99~\textmu m. We find that the shape of the resulting light curve is robust against issues arising from instrumental flexure, as tested using calibration frames collected throughout the observation sequence. Using a multiple linear regression approach with eight decorrelation parameters, we modelled and removed non-shared trends from the differential white-light curve. We find that airmass and air temperature are the most correlated parameters with the light curve. We also analyse publicly available data from the TESS mission to check for variability in the host star HD~1160~A, and confirm that it is non-varying at the 0.03\% level.

We find that the detrended differential white-light curve of HD~1160~B shows sinusoidal-like variability over a short timescale. By fitting the unbinned light curve with a sinusoid, we identify that the variability has a semi-amplitude of $\sim$8.8\% and a period of $\sim$3.24~hours. When binned to 18~minutes of integration time per bin, we achieve a light curve precision at the 3.7\% level. After thorough investigation and rejection of systematic noise sources, we attribute this variability as likely due to heterogeneous features in the atmosphere of the companion, rotating in and out of view as it rotates. We find that if the period of this variability reflects the rotation period of HD~1160~B, physical limitations suggest that it is rotating at close to its breakup period. Alternatively, the short period variability in the light curve of HD~1160~B may arise from multiple periodic features in its atmosphere with different phase offsets. Furthermore, light curves in the 30 individual wavelength channels in the 3.59-3.99~\textmu m range show tentative evidence of an increase in variability amplitude at longer wavelengths. Further observations at additional epochs will help to confirm and characterise the variability of HD~1160~B and to determine its physical explanation. 

The precision that we achieve in the detrended differential white-light curve is the greatest achieved from ground-based studies of sub-arcsecond high-contrast companions to date. However, direct comparisons to other ground-based studies that instead use satellite spots to search for variability in the light curves of high-contrast companions are not straightforward due to the different magnitude and contrast of the observed systems, with HD~1160~B having a more favourable contrast \citep{2016ApJ...820...40A, 2021MNRAS.503..743B, 2022AJ....164..143W}. The RMS of the detrended differential light curve for HD~1160~B as a function of bin size follows the same trend as the theoretical white noise expectation with no evidence of yet approaching a noise floor. This indicated that the single night of data analysed here is not yet systematic-limited, and that further observations from additional epochs could enable greater sensitivity to be reached. A deeper investigation of this type of data and its precision, including injection-recovery tests to test how effectively known variability signals can be recovered and which systematics have the greatest impact, is forthcoming (Sutlieff et al., in preparation).

While JWST will measure the variability of fainter, lower mass companions from space with unprecedented precision, its comparatively smaller aperture means it cannot outperform the largest AO-equipped ground-based telescopes at  separations $\lesssim0.5\arcsec$ at $\sim$3.5~\textmu m \citep{2022SPIE12180E..3QG}, so companions such as Jupiter analogues at the closest angular separations, for all but the nearest stars, remain accessible only to ground-based monitoring techniques for the coming decade. Ground-based differential spectrophotometry with the vAPP is therefore highly complementary to space-based approaches for measuring the variability of high-contrast exoplanet and brown dwarf companions, and for searching for their transiting exomoons or binary planets. Moreover, ground-based telescopes can reach much higher spectral resolution, which then enables line profile variability studies to map atmospheric features, including storms and hurricanes like Jupiter's Great Red Spot, via Doppler imaging. These results are promising for further variability studies using vAPP coronagraphs on current and upcoming instruments and telescopes, which include ERIS on the VLT and METIS on the ELT.

\section*{Acknowledgements}\label{ack}
The authors would like to thank Frank Backs and Elisabeth C. Matthews for valuable discussions that improved this work. We also thank our anonymous referee whose comments helped us to improve this manuscript. BJS is fully supported by the Netherlands Research School for Astronomy (NOVA). JLB acknowledges funding from the European Research Council (ERC) under the European Union’s Horizon 2020 research and innovation program under grant agreement No 805445. This paper is based on work funded by the United States National Science Foundation (NSF) grants 1608834, 1614320, and 1614492. The research of DD and FS leading to these results has received funding from the European Research Council under ERC Starting Grant agreement 678194 (FALCONER).

We acknowledge the use of the Large Binocular Telescope Interferometer (LBTI) and the support from the LBTI team, specifically from Emily Mailhot, Jared Carlson, Jennifer Power, Phil Hinz, Michael Skrutskie, Travis Barman, Ilya Ilyin, and Ji Wang. The LBT is an international collaboration among institutions in the United States, Italy and Germany. LBT Corporation partners are: The University of Arizona on behalf of the Arizona Board of Regents; Istituto Nazionale di Astrofisica, Italy; LBT Beteiligungsgesellschaft, Germany, representing the Max-Planck Society, The Leibniz Institute for Astrophysics Potsdam, and Heidelberg University; The Ohio State University, representing OSU, University of Notre Dame, University of Minnesota and University of Virginia. We gratefully acknowledge the use of Native land for our observations. LBT observations were conducted on the stolen land of the Ndee/Nnēē, Chiricahua, Mescalero, and San Carlos Apache tribes.

This publication makes use of data products from the Two Micron All Sky Survey, which is a joint project of the University of Massachusetts and the Infrared Processing and Analysis Center/California Institute of Technology, funded by the National Aeronautics and Space Administration and the National Science Foundation. This work has made use of data from the European Space Agency (ESA) mission
{\it Gaia} (\url{https://www.cosmos.esa.int/gaia}), processed by the {\it Gaia}
Data Processing and Analysis Consortium (DPAC, \url{https://www.cosmos.esa.int/web/gaia/dpac/consortium}). Funding for the DPAC
has been provided by national institutions, in particular the institutions participating in the {\it Gaia} Multilateral Agreement. This paper includes data collected with the TESS mission, obtained from the Mikulski Archive for Space Telescopes (MAST) data archive at the Space Telescope Science Institute (STScI). Funding for the TESS mission is provided by the NASA Explorer Program. STScI is operated by the Association of Universities for Research in Astronomy, Inc., under NASA contract NAS 5–26555. Support for MAST for non-HST data is provided by the NASA Office of Space Science via grant NNX13AC07G and by other grants and contracts. This research has made use of the NASA Exoplanet Archive, which is operated by the California Institute of Technology, under contract with the National Aeronautics and Space Administration under the Exoplanet Exploration Program. This research has made use of NASA’s Astrophysics Data System. This research has made use of the SIMBAD database, operated at CDS, Strasbourg, France \citep{2000A&AS..143....9W}. This research made use of SAOImageDS9, a tool for data visualization supported by the Chandra X-ray Science Center (CXC) and the High Energy Astrophysics Science Archive Center (HEASARC) with support from the JWST Mission office at the Space Telescope Science Institute for 3D visualization \citep{2003ASPC..295..489J}. This work made use of the whereistheplanet\footnote{\url{http://whereistheplanet.com/}} prediction tool \citep{2021ascl.soft01003W}. This work makes use of the Python programming language\footnote{Python Software Foundation; \url{https://www.python.org/}}, in particular packages including NumPy \citep{harris2020array}, Astropy \citep{astropy:2013, astropy:2018}, SciPy \citep{2020SciPy-NMeth}, HCIPy \citep{por2018hcipy}, PynPoint \citep{2012MNRAS.427..948A, 2019A&A...621A..59S}, scikit-image \citep{scikit-image}, scikit-learn \citep{scikit-learn}, statsmodels \citep{seabold2010statsmodels}, pandas \citep{mckinney-proc-scipy-2010, jeff_reback_2022_6408044}, Photutils \citep{larry_bradley_2022_6385735}, and Matplotlib \citep{Hunter:2007}.
\section*{Data Availability}\label{ava}
The data from the LBT observations underlying this article are available in the Research Data Management Zenodo repository of the Anton Pannekoek Institute for Astronomy, at \url{https://doi.org/10.5281/zenodo.5617572}.



\bibliographystyle{mnras}
\bibliography{bibliography} 








\bsp	
\label{lastpage}
\end{document}